\documentclass[onecolumn,floatfix,superscriptaddress,a4paper,
               showpacs,showkeys,nofootinbib,preprint]{revtex4}

\usepackage{epsfig}
\usepackage{amssymb,latexsym,amsmath}
\newcommand{\eq}[1]{\begin{align} #1 \end{align}}

\begin{document}

\title{Semi-Inclusive Distributions in Statistical Models}

 \author{V. V. Begun}
 \affiliation{Bogolyubov Institute for Theoretical Physics, Kiev, Ukraine}
 \affiliation{Frankfurt Institute for Advanced Studies, Frankfurt, Germany}

 \author{M. Ga\'zdzicki}
 \affiliation{Institut f\"ur Kernphysik, University of Frankfurt, Frankfurt, Germany}
 \affiliation{Jan Kochanowski University, Kielce, Poland}

 \author{M. I. Gorenstein}
 \affiliation{Bogolyubov Institute for Theoretical Physics, Kiev, Ukraine}
 \affiliation{Frankfurt Institute for Advanced Studies, Frankfurt, Germany}

%-----------------------------------------------------------
\begin{abstract}

The semi-inclusive properties of the system of neutral and charged
particles with net charge equal to zero are considered in the
grand canonical, canonical and micro-canonical ensembles as well
as in micro-canonical ensemble with scaling volume fluctuations.
Distributions of neutral particle multiplicity and charged
particle momentum are calculated as a function of the number of
charged particles. Different statistical ensembles lead to
qualitatively different dependencies. They are being compared with
the corresponding experimental data on multi-hadron production in
$p+p$ interactions at high energies.
%Qualitative disagreement is found for the grand
%canonical,
%canonical, and micro-canonical ensembles. %Only
%{\bf Whereas} the results of the micro-canonical ensemble with
%scaling volume fluctuations follow the trends  observed in the
%data.

\end{abstract}



\maketitle

%-----------------------------------------------------------------------
%

%
%%%%%%%%%%%%%%%%%%%%%%%%%%%%%%%%%%%%%%%%%%%%%%%%%%%%%%%%%%%%%%%%%%%%%%%%%%
%
\section{Introduction}\label{S-I}
%
%%%%%%%%%%%%%%%%%%%%%%%%%%%%%%%%%%%%%%%%%%%%%%%%%%%%%%%%%%%%%%%%%%%%%%%%%%
%
In relativistic high-energy collisions many properties of produced
particles follow simple rules of statistical mechanics. The single
particle momentum spectrum approximately has the Boltzmann form,
$dN/d^3p\sim\exp[-(p^2+m^2)^{1/2}/T]$, in the local rest frame of
produced matter~\cite{Ha:65}. The mean particle multiplicity of
heavy particles ($m\gg T$) is also governed by the Boltzmann factor,
$\langle N\rangle\sim\exp(-m/T)$. Here $T$, $p$, and $m$ are the
temperature parameter, the particle momentum, and its mass,
respectively. The temperature parameter  extracted from the data
on $p+p$ interactions is in the range of 160-190~MeV~\cite{Be:97}.
Thus, almost all particles are produced at low transverse momenta,
$p_T$, and with low masses ($p_T,\,m\leq 2$ GeV).

However, the standard statistical approach fails to reproduce the
KNO scaling \cite{kno} of multiplicity distributions observed in
the data on $p+p$, $p+\overline{p}$, and $e^++e^-$ collisions
\cite{knog1,knog2,knog3}. The other problems are a power law
behavior of the single particle transverse momentum
spectrum at large transverse momenta, $dN/d^3p\sim p_T^{-K_p}$, and
a power law dependence of a mean multiplicity of heavy particles,
$\langle N\rangle\sim m^{-K_m}$, where $K_m\cong K_p-3$
\cite{powerlaw}. In our previous paper \cite{powerlaw2} an
extension of the standard statistical approach was suggested to a
region of large transverse momenta and/or large particle masses
($p_T,m\ge 3$~GeV) by taking into account volume fluctuations. The
proposed model, the statistical ensemble with scaling volume
fluctuations (MCE/sVF), allows to solve the above mentioned
problems of the statistical approach.

Hadron production in high-energy collisions is characterized by
two types of quantities: inclusive and semi-inclusive ones.
Statistical models are usually used to describe inclusive
quantities like mean multiplicity or mean transverse momentum.
They are calculated by the summation over all microstates of the
system with corresponding statistical weights. In the present
paper we study selected properties of semi-inclusive quantities
within statistical models. In this case the statistical summation
is restricted by the additional condition, e.g. by a requirement
that charged hadron multiplicity is fixed. The grand canonical,
canonical and micro-canonical ensembles as well as MCE/sVF will be
used. In particular, the mean multiplicity of neutral particles
and average transverse momentum of charged particles are
considered at a fixed charged particle multiplicity. The obtained
model predictions are compared to the trends observed in the
experimental data.

For the sake of simplicity, the system of non-interacting massless
Boltzmann particles -- neutral, positively charged and negatively
charged -- with the total net charge equal to zero $Q=N_+-N_-=0$
is considered. The degeneracy factors are assumed to be
$g_0=g_+=g_-=1$ and the temperature parameter is set to $T =
160$~MeV for quantitative calculations.

The paper is organized as follows. The joint multiplicity
distributions of neutral and negatively charged particles,
correlations, semi-inclusive averages, and the effects of quantum
statistic are calculated in Section II. Semi-inclusive momentum
spectra are obtained and discussed in Section III. A comparison
with available data is presented  in Section IV. Summary presented
in Section V closes the paper.

\section{multiplicity distributions }
%

%
%%%%%%%%%%%%%%%%%%%%%%%%%%%%%%%%%%%%%%%%%%%%%%%%%%%%%%%%%%%%%%%%%%%%%%%%%%
%
\subsection{Grand Canonical Ensemble}
%
%%%%%%%%%%%%%%%%%%%%%%%%%%%%%%%%%%%%%%%%%%%%%%%%%%%%%%%%%%%%%%%%%%%%%%%%%%
%
The grand canonical ensemble (GCE) is defined by the system volume $V$,
temperature $T$,  and
charge chemical potential $\mu_Q$.
The  chemical potential $\mu_Q$ regulates an average
value of the conserved charge $Q$.  For the system with zero
net charge considered here  $\mu_Q$ is equal to zero.
The mean particle multiplicities and the average energy in the GCE are:
 \eq{\label{Ngce}
\langle N_0\rangle_{gce}
 \;&=\; \langle N_+\rangle_{gce}\;=\; \langle N_-\rangle_{gce}
 \;\equiv\; \overline{N}=V\,T^3/\pi^2\; ,\\
 \langle E\rangle_{gce}
   \;& =\; 3T\,\langle N_0\rangle_{gce} \;+\; 3T\,\langle N_-\rangle_{gce}~+~3T\,\langle N_+\rangle_{gce}
   ~\equiv~\overline{E} \;=\; 9T\,\overline{N}~.\label{Egce}
 }
In the GCE the neutral and charged multiplicities $N_0$ and $N_+$,
$N_-$ are uncorrelated and obey the Poisson distribution. Thus,
the joint distribution of  neutral $N_0$ and negatively charged
particles $N_-$ is given by the product of two Poisson
distributions which can be approximated by the product of two
Gauss distributions at $\overline{N}\gg 1$:
 \eq{\label{Pgce}
 P_{gce}(N_0,N_-) \;&=\; \frac{\overline{N}^{N_0}}{N_0!}\exp\left(-\overline{N}\right)
   ~\times ~
 \frac{\overline{N}^{N_-}}{N_-!}\exp\left(-\overline{N}\right) \\\nonumber
% P_{gce}^G(N,N_-)
 & \cong ~(2\pi \overline{N})^{-1/2}
        \exp\left[-\frac{\left(N_0-\overline{N}\right)^2}
        {2\overline{N}}\right]
 ~\times~ (2\pi\overline{N})^{-1/2}~
        \exp\left[-\frac{\left(N_--\overline{N}\right)^2}
        {2\overline{N}}\right]\;.
  }

\subsection{Canonical Ensemble}
The canonical ensemble (CE) is described by the variables $V,T,Q$.
The GCE expressions (\ref{Ngce}-\ref{Egce}) for average quantities
remain valid in the CE at $\overline{N}\gg 1$. From the assumption
$Q = 0$  follows that $N_+=N_-$. Consequently, the distribution of
$N_+$ and $N_-$ in the CE is narrower than in the GCE~\cite{CE}.
The CE distribution of neutral particles remains the same as in
the GCE as it is not constrained by  charge conservation law. The
joint distribution of  neutral  and negatively charged  particles
is given by \cite{CE}:
 \eq{\label{Pce}
 P_{ce}(N_0,N_-) \;&=\;
 \frac{\overline{N}^{N_0}}{N_0!}\exp\left(-\overline{N}\right)
 ~\times ~
 \frac{1}{I_0(2\overline{N})}\;\frac{\overline{N}^{2N_-}}{(N_-!)^2} \\\nonumber
 %}
%
 & \cong ~(2\pi \overline{N})^{-1/2}
        \exp\left[-\frac{\left(N_0-\overline{N}\right)^2}
        {2\overline{N}}\right] ~\times~ (\pi\overline{N})^{-1/2}~
        \exp\left[-\frac{\left(N_--\overline{N}\right)^2}
        {\overline{N}}\right]~,
 }
where $I_0$ is the modified Bessel function.

\subsection{Micro-Canonical Ensemble}
The micro-canonical ensemble (MCE) is described by the variables
$V,E,Q$.
The MCE partition function for $N_0$ neutral and $N_+=N_-$
positively and negatively charged  massless particles reads~\cite{mce2}:
 \eq{\label{Omega}
 \Omega_{N_0,N_-}(E,\,V)
 \;=\; \frac{1}{N_0!}\, \frac{1}{(N_-)!^2}\,
        \left(\frac{V}{\pi^2}\right)^{N_0+2N_-}
%        \left(\frac{g_-V}{\pi^2}\right)^{2N_-}
  \,\frac{E^{3N_0+6N_--1}}{\Gamma(3N_0+6N_-)}\, ,
 }
where $\Gamma$ is the Euler gamma function.
The joint probability
distribution of  $N_0$ and $N_-$ in the MCE  is,
 \eq{\label{Pmce}
 P_{mce}(N_0,N_-)~=~\frac{\Omega_{N_0,N_-}(E,V)}{\Omega(E,V)}~,
 }
where
% \eq{\label{MCE-PF}
 $\Omega(E,V)=\sum_{N_0,N_-}\Omega_{N_0,N_-}(E,V)$.
%
%
%
%The mean value of any  quantity $X$ can be expressed as an average
%over the corresponding multiplicity distribution:
%
%\eq{\label{incl-av}
%
%\langle X \rangle~=~\sum_{N_0,N_-} X(N_0,N_-)~P(N_0,N_-)~.
%
%}
%

\subsection{Average Multiplicities, Fluctuations and Correlations}

The mean quantities in different statistical ensembles can be
expressed as
\eq{\label{incl-av}
 \langle X \rangle
% \;\equiv\; \overline{X}
~=\sum_{N_0,N_-} X(N_0,N_-)~P(N_0,N_-)~.
 }
For the MCE distribution (\ref{Pmce}) one obtains:
 \eq{\label{Nmce}
 \langle N_0\rangle_{mce} \;\cong\; \langle N_-\rangle_{mce}
 \;\cong\; \frac{1}{3\sqrt{3\pi}}~\left(VE^3\right)^{1/4}.
 }
If the MCE energy equals  the average energy of the GCE and CE,
$E=\overline{E}$, then the average MCE multiplicities (\ref{Nmce})
become  equal to those in the GCE and CE (\ref{Ngce}). This
reflects the equivalence of the GCE, CE, and MCE in the
thermodynamic limit.
%\footnote{Corrections to (\ref{Nmce}) are
%already small at $\overline{N}\ge 3$,
%Practically the ensembles are almost equivalent for the mean
%values already for $\langle N\rangle_{ce}>5$ and for $\langle
%N\rangle_{mce}>2$, see \cite{CE}},
%see Ref.~\cite{mce1}.} $\overline{N}\gg 1$.
%
However, the multiplicity distributions  are different in these
ensembles even in the thermodynamic limit. As shown in the
Appendix~A the MCE distribution (\ref{Pmce}) for $\overline{N} \gg
1$ can be approximated as
 \eq{\label{PmceG}
 P_{mce}(N_0,N_-)
 \;\cong\; \frac{\sqrt{2}}{\pi\,\overline{N}}\;
       \exp\left[\; -\,\frac{(N_0-\overline{N})^2}{\overline{N}}
 \;-\; \frac{2(N_0-\overline{N})(N_--\overline{N})}{\overline{N}}
 \;-\; \frac{3(N_--\overline{N})^2}{\overline{N}}
 \right]~.
 }
The distributions $P(N_0,N_-)$ in the GCE (\ref{Pgce}),  CE
(\ref{Pce}), and MCE (\ref{Pmce}) can be written in a general form
of the bivariate normal distribution,
%
%For further study it is convenient to rewrite the two dimensional
%distribution (\ref{Pgce}) in the form of bivariate Gauss
%distribution:
%
 \eq{\label{P-bivar}
 & P(N_0,N_-)
 \;=\; \frac{1}{2\pi\,\overline{N}\,\sqrt{\omega^0\cdot\omega^-(1-\rho^2)}}\;
         \nonumber \\
 & \exp\left[\; -\;\frac{1}{2\,\overline{N}\,(1-\rho^2)}
 \left(\frac{(N_0-\overline{N})^2}{\omega^0}
 \;-\; 2\,\rho\;\frac{(N_0-\overline{N})(N_--\overline{N})}
        {\sqrt{\omega^0\cdot\omega^-}}
 \;+\; \frac{(N_--\overline{N})^2}{\omega^-} \right)\right]\;,
 }
where $\omega^0$ and $\omega^-$ are the scaled variances defined
as:
 \eq{\label{omega}
 \omega^0 ~\equiv~ \frac{\langle N_0^2\rangle~-~\langle
        N_0\rangle^2}{\langle N_0\rangle}~,\qquad
 \omega^- ~\equiv~ \frac{\langle N_-^2\rangle~-~\langle
        N_-\rangle^2}{\langle N_-\rangle}~,
 }
and $\rho$ is the correlation coefficient:
 \eq{\label{Rho}
 \rho~\equiv ~\rho^{0-} \;\equiv\; \frac{\langle N_0\,N_-\rangle
         \;-\; \langle N_0\rangle\,\langle N_-\rangle}
         {\sqrt{[\,\langle N_0^2\rangle\;-\; \langle N_0\rangle^2\,]
 \cdot    [\,\langle N_-^2\rangle \;-\; \langle N_-\rangle^2\,]}}
 \;=\; \frac{\langle N_0\,N_-\rangle
 \;-\;      \langle N_0\rangle\,\langle N_-\rangle}
                    {\sqrt{\omega^0\cdot\langle N_0\rangle
                    \cdot\omega^-\cdot\langle N_-\rangle}}\;.
 }
%
%It is zero if there is no correlation and $\rho=1$ if one has
%absolute correlation when the distribution (\ref{P-bivar}) become
%the delta function $\delta(N_0-N_-)$.
The averaging in  Eqs.~(\ref{omega},\ref{Rho}) is expressed
according to Eq.~(\ref{incl-av}).
% over the multiplicity
%distribution (\ref{Pmce}), where $\langle N^2\rangle$ is
%calculated according to Eq.~(\ref{incl-av}) with $X = N^2$ and
%$\langle N_0N_-\rangle$ with $X = N_0N_-$.
%

The scaled variances in the GCE correspond to the uncorrelated
Poisson distributions (\ref{Pgce}):
 \eq{\label{w-gce-Boltz}
 \omega^0_{gce} \;=\; \omega^-_{gce} \;=\; \omega^+_{gce} \;=\; 1\;,
 }
and the correlation coefficient (\ref{Rho}) is obviously equal to
zero. One can similarly introduce the coefficients $\rho^{0+}$ and
$\rho^{+-}$. They are also equal to zero in the GCE.
%and appear if one, for example, impose global conservation
%laws on the system.
%
%

In the CE,
 \eq{\label{w-ce-Boltz}
 \omega^0_{ce}
 %\;=\; \omega^0_{gce}
 \;=\; 1\;,
 %\qquad
 %\text{but}
 \qquad
 \omega^-_{ce} \;=\;\omega^+_{ce} \;=\; \frac{1}{2}\;,
 }
for the distribution (\ref{Pce}).
%Thus, despite of equivalence of
%the GCE and CE, the multiplicity distributions are different there
%even in the thermodynamic limit \cite{CE}.
%The scaled variance for
%positive particles is equal to negative one,
%$\omega^+_{ce}\equiv\omega^-_{ce}$, because of $N_+=N_-$.
%
%The correlation coefficients $\rho^{0-}_{ce}$ and $\rho^{0+}_{ce}$
%are zero as in the GCE. However,
The strong correlation, $N_+=N_-$, in each microscopic state of
the CE
%instead of $\langle
%N_+\rangle = \langle N_-\rangle$ in the GCE,
leads to the largest possible value of the correlation
coefficient:
 \eq{\label{Rhopm}
 \rho^{+-}_{ce}
  \;=\; \frac{\langle N_+\,N_-\rangle_{ce}
  \;-\;      \langle N_+\rangle_{ce}\,\langle N_-\rangle_{ce}}
                    {\sqrt{\omega^+_{ce}\cdot\langle N_+\rangle_{ce}
                    \cdot\omega^-_{ce}\cdot\langle N_-\rangle_{ce}}}
  \;=\; 1\;.
 }
However, similar to the GCE, there are no correlations between
neutral and charged particles, $ \rho^{0\pm}_{ce}=0$.

The scaled variances in the MCE are:
 \eq{\label{w-mce-Boltz}
 \omega^0_{mce} \;=\; \frac{3}{4}\;,
 %\qquad
 %\text{and}
 \qquad
 \omega^-_{mce} \;=\; \omega^+_{mce} \;=\;\frac{1}{4}\;.
 }
They reflect the suppression of fluctuations of neutral particles
in the MCE comparing to the GCE and CE, and stronger suppression
of fluctuations of charged particles comparing to the CE. The
correlation coefficient,~
%
% \eq{
 $\rho^{+-}_{mce} \;=\; 1\;,$
% }
%
is the same as in the CE (\ref{Rhopm}).
%, thus,
%$\omega^+_{mce}=\omega^-_{mce}$.
The exact energy conservation in
the MCE leads to a rather strong anti-correlation between neutral
and charged particles:
 \eq{\label{Rho0m}
 \rho^{0-}_{mce} \;=\; \rho^{0+}_{mce} \;=\; -~\frac{1}{\sqrt{3}} \; \cong \; -~0.577\;.
 }
%

%The multiplicity fluctuations are often characterized by
%the scaled variance
%defined as:
%
% \eq{\label{omega}
%
%\omega~
%  \equiv ~\frac{\langle N^2\rangle~-~\langle
%        N\rangle^2}{\langle N\rangle}~,
% }
 %
%where $\langle N^2\rangle$ is calculated according to
%Eq.~(\ref{incl-av}) with $X = N^2$.

\begin{figure}[ht!]
\epsfig{file=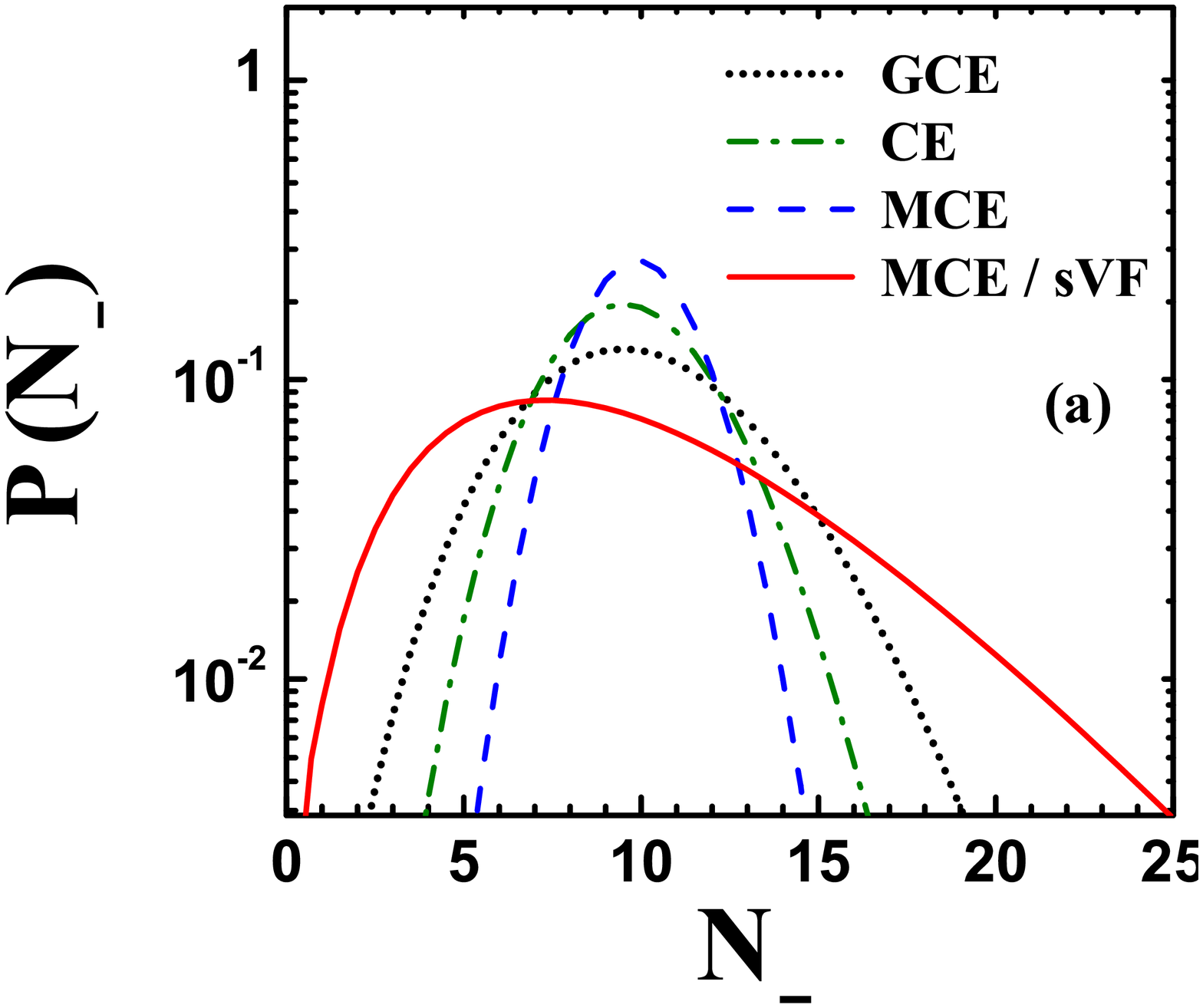,width=0.49\textwidth}\;\;
\epsfig{file=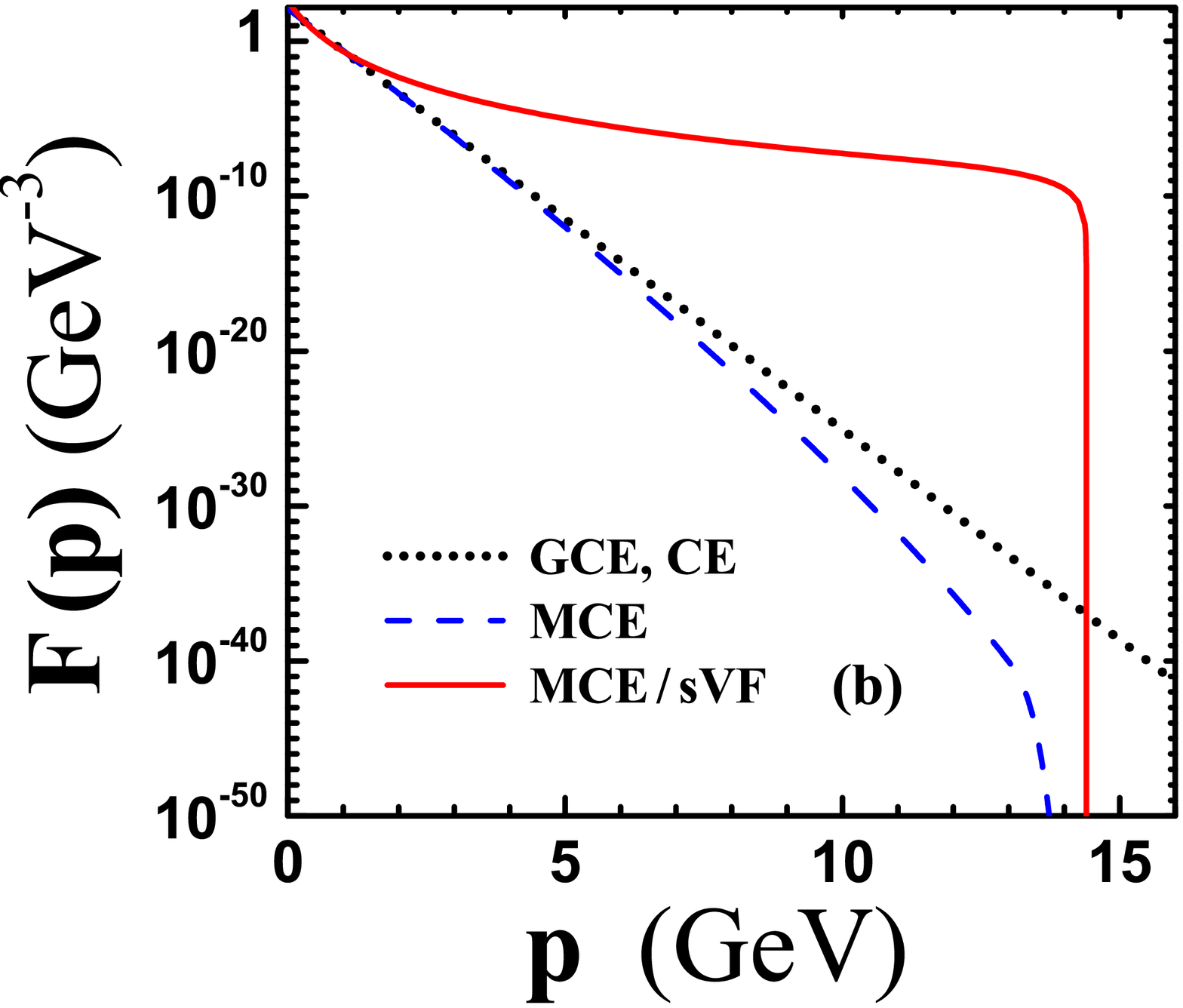,width=0.49\textwidth}
\caption{(Color online)
Examples of the multiplicity distributions ({\it left})
and the inclusive momentum spectra ({\it right}) of negatively charged particles
obtained within the GCE, CE, MCE and MCE/sVF.
The distributions are calculated assuming $\overline{N}=10$ and $T=160$~MeV (see text for details).
 } \label{fig-in}
\end{figure}

In the large volume limit the multiplicity distribution of
negatively charged particles in the GCE, CE, and MCE can be
approximated by normal distribution \cite{clt}:
\eq{\label{gauss}
P(N_-)~=~\sum_{N_0}P(N_-,N_0)~\cong~ (2\pi \omega^-\overline{N})^{-1/2}~
        \exp\left[-~\frac{\left(N_--\overline{N}\right)^2}
        {2\omega^-\overline{N}}\right]\;,
}
%
%The scaled variance $\omega^-$ given by Eq.~(\ref{gauss}) equals
with $\omega_{gce}^-=1$, $\omega_{ce}^-=1/2$, and
$\omega^{-}_{mce}=1/4$ in the GCE, CE and MCE ensembles,
respectively.  The $N_-$ distribution in these ensembles is
presented in Fig.~\ref{fig-in}, {\it left}. As the considered
system has zero charge, the distributions $P(N_+)$ are equal to
$P(N_-)$ ones in all statistical ensembles.

The  neutral particle multiplicity distribution reads:
\eq{\label{gauss1}
P(N_0)~=~\sum_{N_-}P(N_-,N_0)~\cong~ (2\pi \omega^0\overline{N})^{-1/2}~
        \exp\left[-~\frac{\left(N_0-\overline{N}\right)^2}
        {2\omega^0\overline{N}}\right]\;,
}
%
%where the scaled variance $\omega^0$ in different ensembles is:
with $\omega_{gce}^0=\omega_{ce}^0=1$ and $\omega^{0}_{mce}=3/4$.
%The exact charge and energy conservation lead therefore to a
%suppression of the multiplicity fluctuations\footnote{
% There is
%more space for fluctuation in the MCE of both neutral and charged
%particles.
%Note that the MCE with charged particles only gives
%$\omega_{mce}^-=1/8$, and with neutral particles only
%the
%corresponding scaled variances equal
%$\omega_{mce}^0=1/4$, see Ref.~\cite{mce1}.}.

\subsection{ MCE with scaling Volume Fluctuations}\label{S-III}

The MCE with scaling volume fluctuations (MCE/sVF)
\cite{powerlaw2} is described by the variables $E$, $Q$ and $V$ as
well as the distribution function defining the scaling volume
fluctuations\footnote{Statistical ensembles with fluctuating
extensive quantities are discussed in recent papers
\cite{alpha,volume}.}. All quantities calculated within the
MCE/sVF will be denoted by the subscript $\alpha$. For a
description of the volume fluctuations it is convenient to
introduce an auxiliary variable $y$ as:
 \eq{\label{Vy}
  y \equiv (V/\overline{V})^{1/4}~,
 }
and describe the scaling volume fluctuations
by the
scaling function $\psi_{\alpha}(y)$ (see
Ref.~\cite{powerlaw2} for details).
Experimental data on the multiplicity distribution of charged
hadrons in $p+p$ interactions suggest a simple analytical form of
the $\psi_{\alpha}(y)$ function ~\cite{powerlaw2,stas,stas1}:%KNO-meaning
 \eq{\label{psi-a}
 \psi_{\alpha}(y) \;=\;
 \frac{k^k}{\Gamma(k)}\,y^{k-1}\exp(-k\,y)\;,
 }
with $k = 4$ and
$\Gamma(k)$ being the Euler gamma function.

The joint $N_0$ and $N_-$
distribution in the MCE/sVF equals to:
\eq{\label{Pa}
 P_{\alpha}(N_0,N_-)
 ~=~\int_0^{\infty}dy~P_{mce}(N_0,N_-)~\psi_{\alpha}(y)~,
} where  $P_{mce}(N_0,N_-)$ is given by Eq.~(\ref{Pmce}). The
analytical approximations for $ P_{\alpha}(N_0,N_-)$ are discussed
in Appendix \ref{app-B}.
The inclusive mean multiplicities in the MCE/sVF are:
\eq{\label{Na}
 \langle N_-\rangle_{\alpha}
  ~=\sum_{N_-,N_0}N_-\,P_{\alpha}(N_0,N_-)
  \;\cong\; \overline{N}~,~~~~ \langle
 N_0\rangle_{\alpha}
  ~=\sum_{N_-,N_0}N_0\,P_{\alpha}(N_0,N_-)
  \;\cong\;\overline{N}~,
}
and, thus, they coincide with those in the GCE, CE, and MCE at $\overline{N}\gg 1$.
The inclusive multiplicity distributions in the MCE/sVF are:
 \eq{\label{PNa}
 P_{\alpha}(N_-)~&=~\sum_{N_0} P_{\alpha}(N_0,N_-)
 ~\cong~ \frac{1}{\overline{N}}\; \psi_{\alpha}\left(\frac{N_-}
         {\overline{N}}\right)~,
\\
 P_{\alpha}(N_0)~&=~\sum_{N_-} P_{\alpha}(N_0,N_-)
 ~\cong~ \frac{1}{\overline{N}}\; \psi_{\alpha}\left(\frac{N_0}
         {\overline{N}}\right)~.\label{PNa0}
}
The $P_{\alpha}(N_-)$  distribution is shown in Fig.~\ref{fig-in}.
It is significantly broader then
the corresponding  distributions for the GCE, CE, and MCE.
The scaled variance for negatively charged and neutral particles is:
 \eq{\label{w-a}
 \omega_{\alpha}^-
  \;\cong\; \frac{1}{k}\,\overline{N}
 \;+\; \omega_{mce}^- \;,~~~~
 \omega_{\alpha}^{0}
  \;\cong\; \frac{1}{k}\,\overline{N}
 \;+\; \omega_{mce}^{0} \;.
 }
Thus, in the MCE/sVF, because of the scaling volume fluctuations,
the scaled variance increases in proportion to the mean
multiplicity, while the scaled variance is approximately
independent of mean multiplicity, $\omega \approx const $, in the
GCE, CE, and MCE.

\begin{figure}[ht!]
 \begin{center}
 \epsfig{file=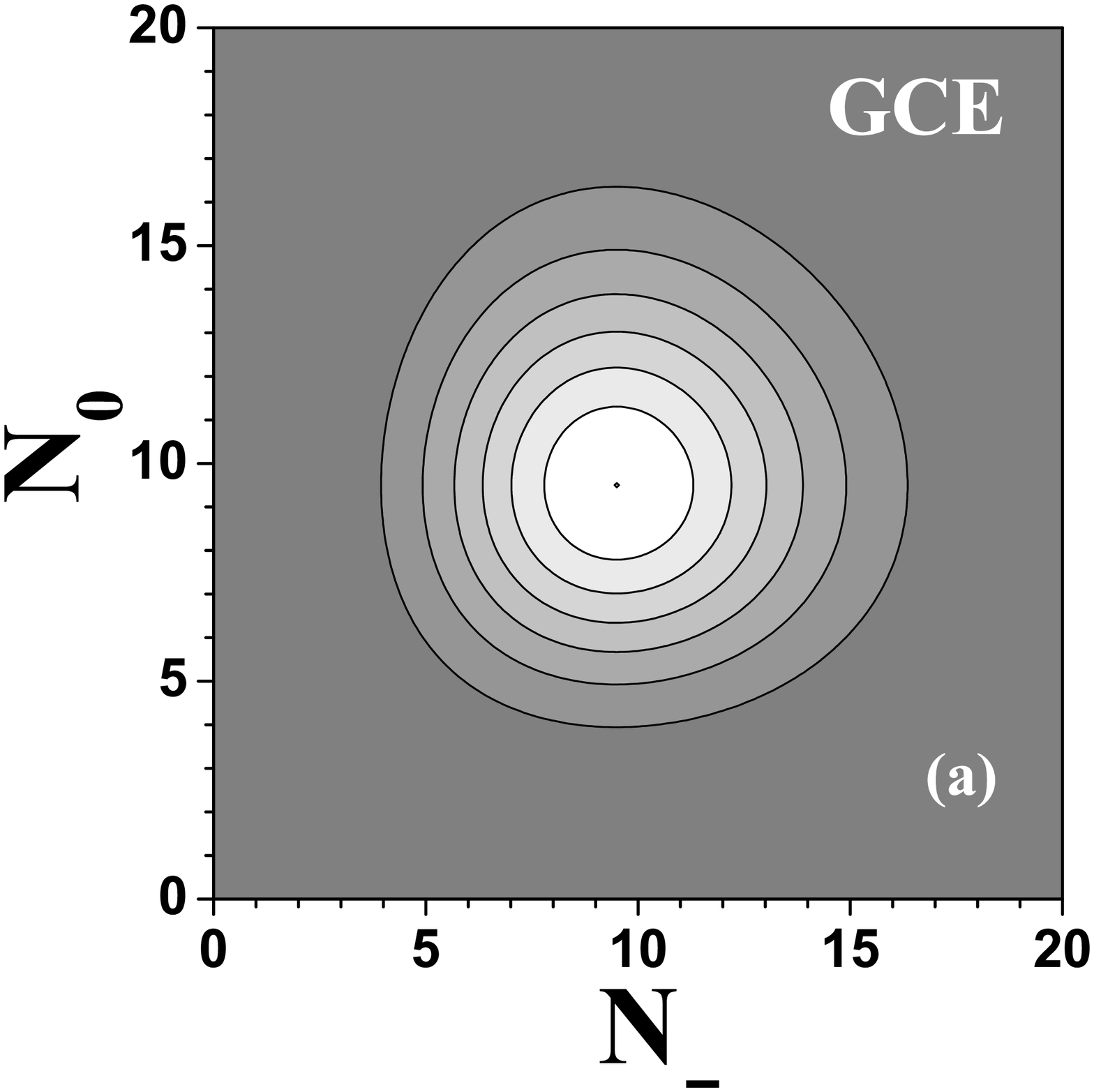,width=0.48\textwidth}\;\;
 \epsfig{file=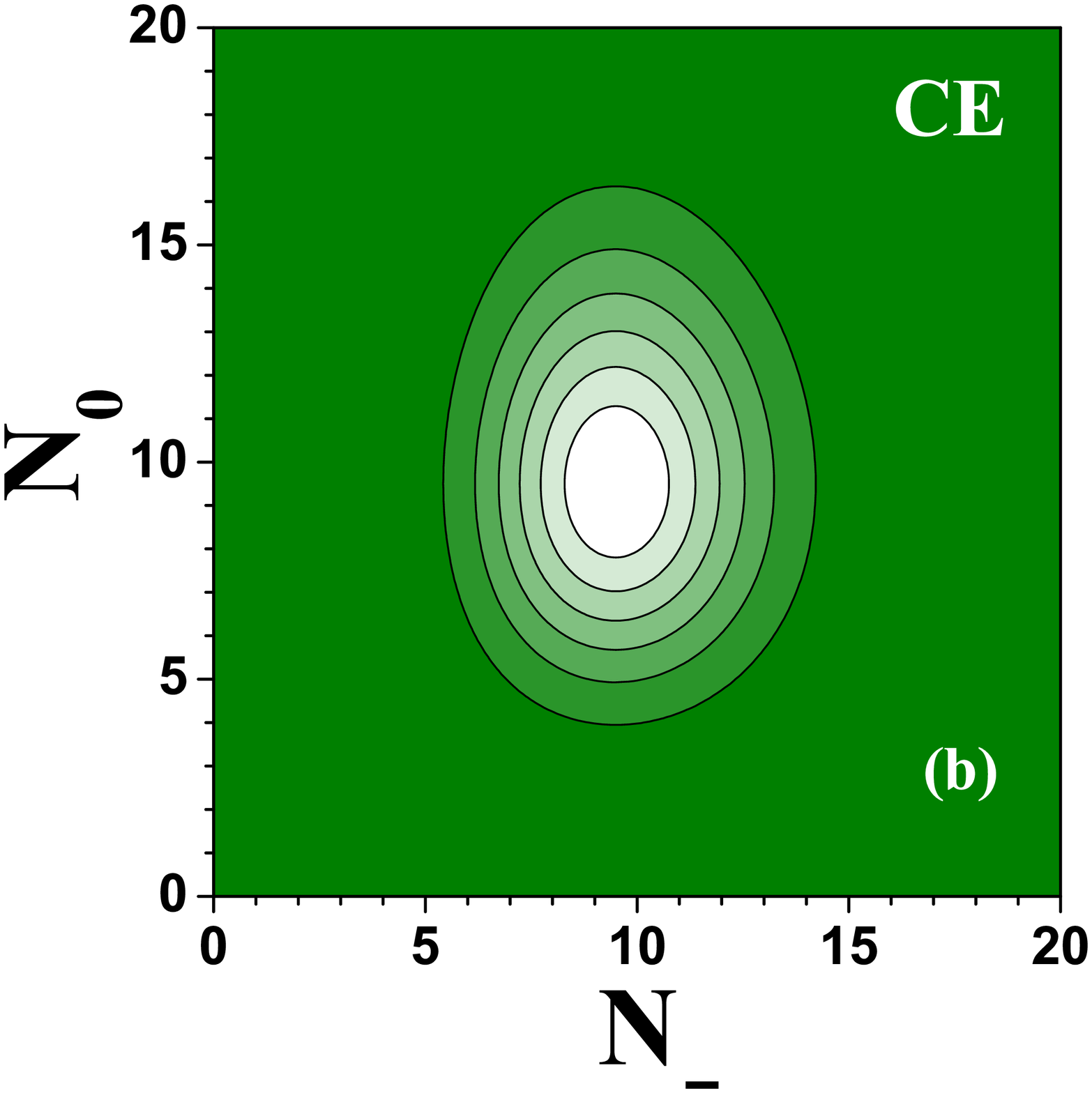,width=0.48\textwidth}\;\;
 \epsfig{file=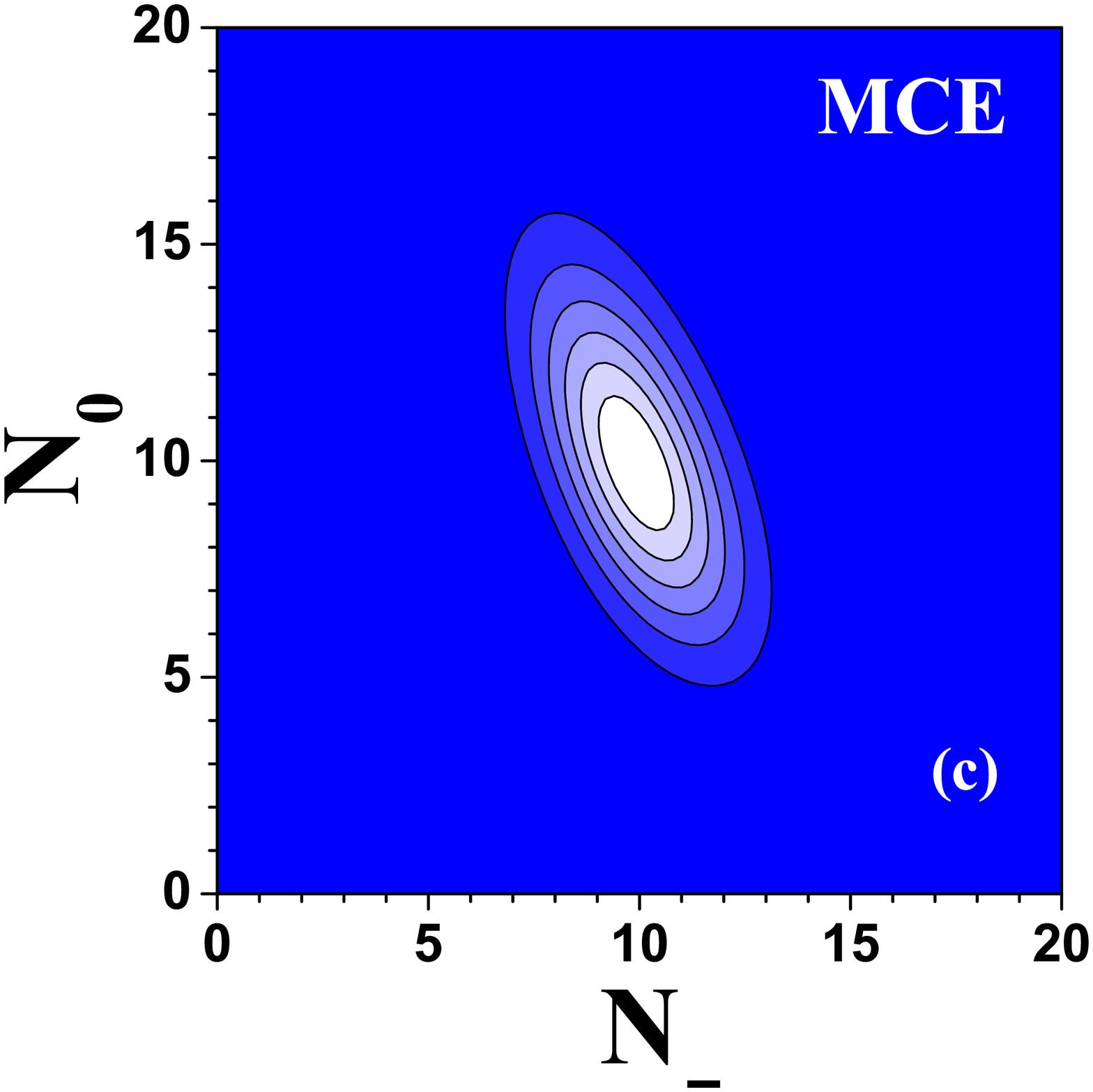,width=0.48\textwidth}\;\;
\epsfig{file=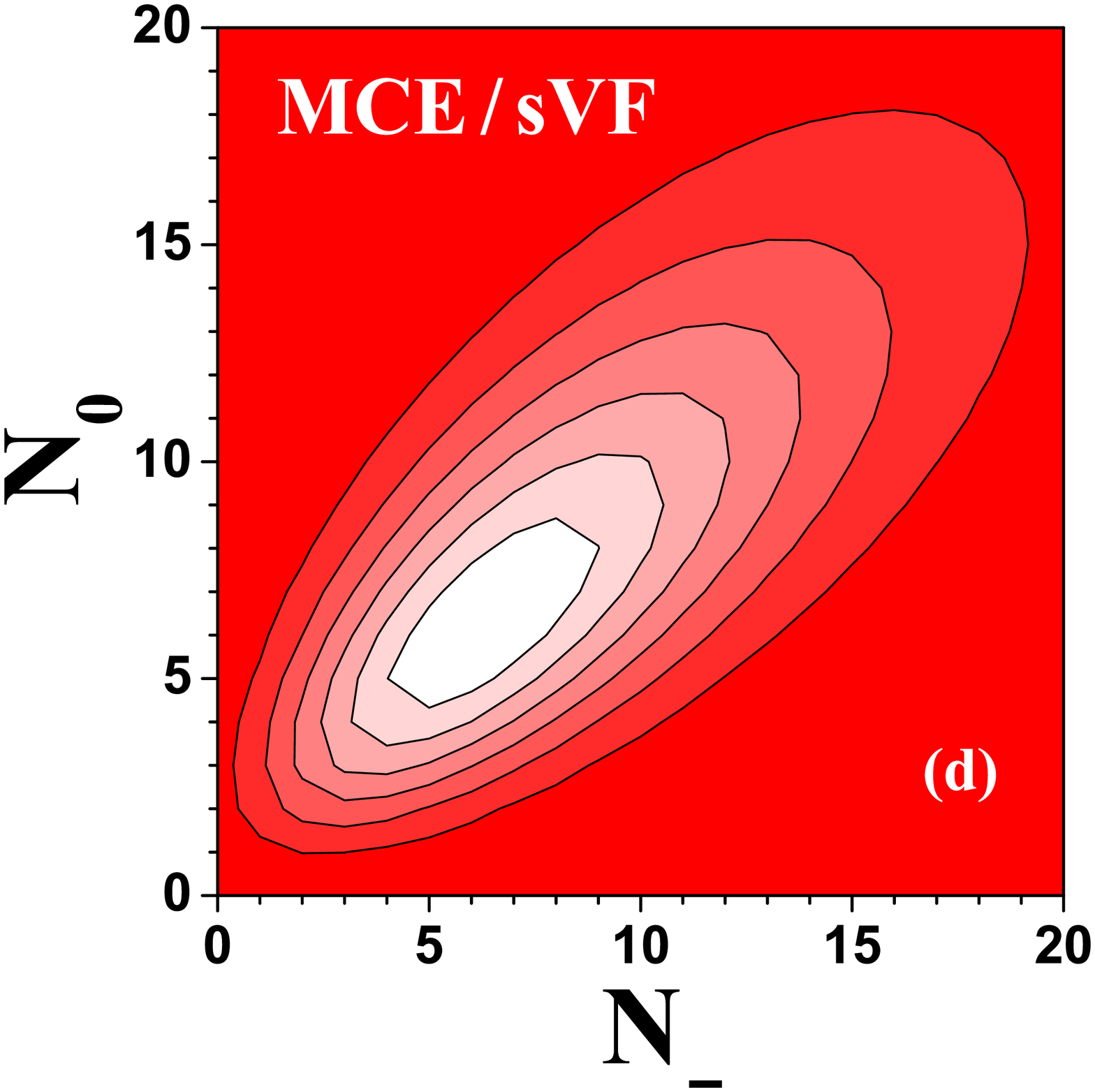,width=0.48\textwidth}\;\;
 \end{center}
 \vspace{-0.7cm}
 \caption{(Color online)
Examples of the joint $N_0$ and
$N_-$ distributions calculated within the GCE ({\it top left}), CE
({\it top right}), MCE ({\it bottom left}) and MCE/sVF ({\it
bottom right}). The distributions are calculated assuming
$\overline{N}=10$ (see text for details).
 \label{fig-P}}
\end{figure}

For illustration of the previously discussed properties
the joint $N_0$ and $N_-$ distributions calculated
within the GCE (\ref{Pgce}), CE (\ref{Pce}),
MCE (\ref{Pmce}), and MCE/sVF (\ref{Pa})  are shown
in~Fig.~\ref{fig-P}.
The multiplicities of neutral and negatively charged particles are
uncorrelated in the GCE and CE (see Fig.~\ref{fig-P} top panels).
They are anti-correlated and correlated in the MCE and MCE/sVF,
respectively. A positive correlation between $N_0$ and $N_-$ in
the MCE/sVF is caused by the scaling volume fluctuations. Note
that
%{\bf The dissymmetry of the distributions (\ref{Pgce}),
%(\ref{Pce}), and (\ref{Pmce}) is caused by
finite size effects are  seen in Fig.~\ref{fig-P}.
%as, for example, the
%Poisson distribution in the GCE (\ref{Pgce}) is not symmetric yet
%for
For example, the Poisson distribution (\ref{Pgce}) is significantly
asymmetric for large deviations from $\overline{N}$.

\begin{figure}[h!]
 \begin{center}
 \epsfig{file=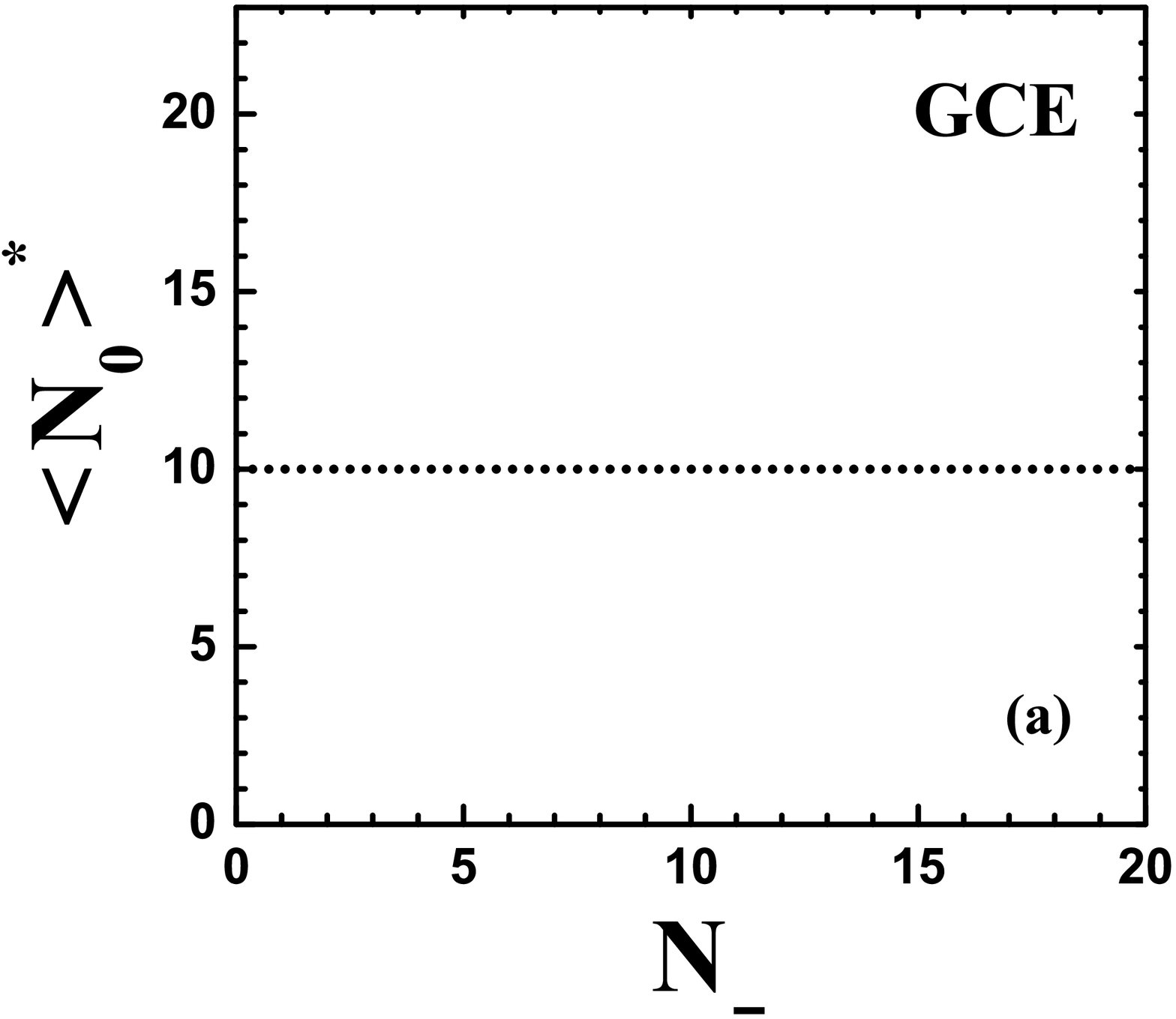,width=0.48\textwidth}
 \epsfig{file=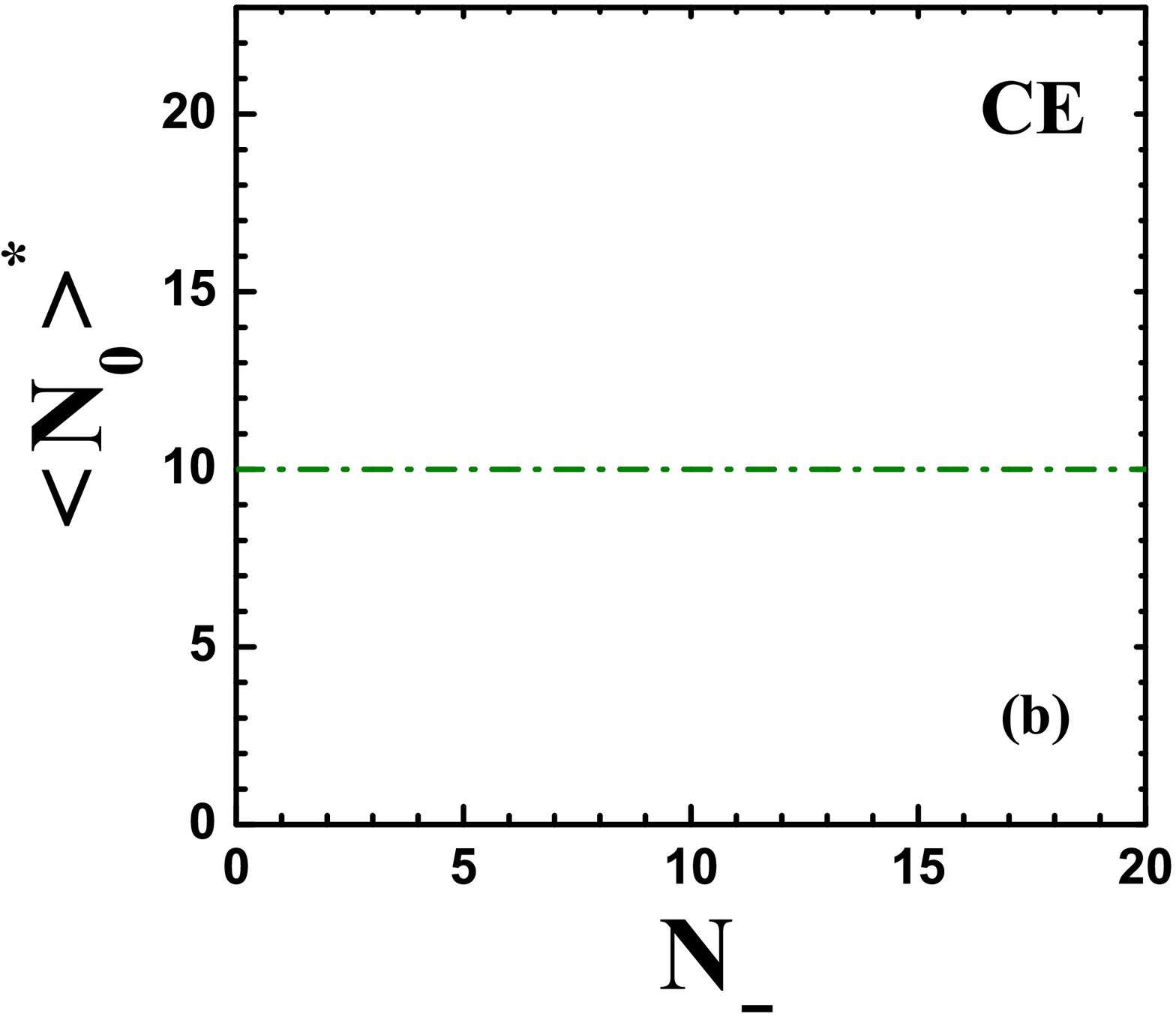,width=0.48\textwidth}
 \epsfig{file=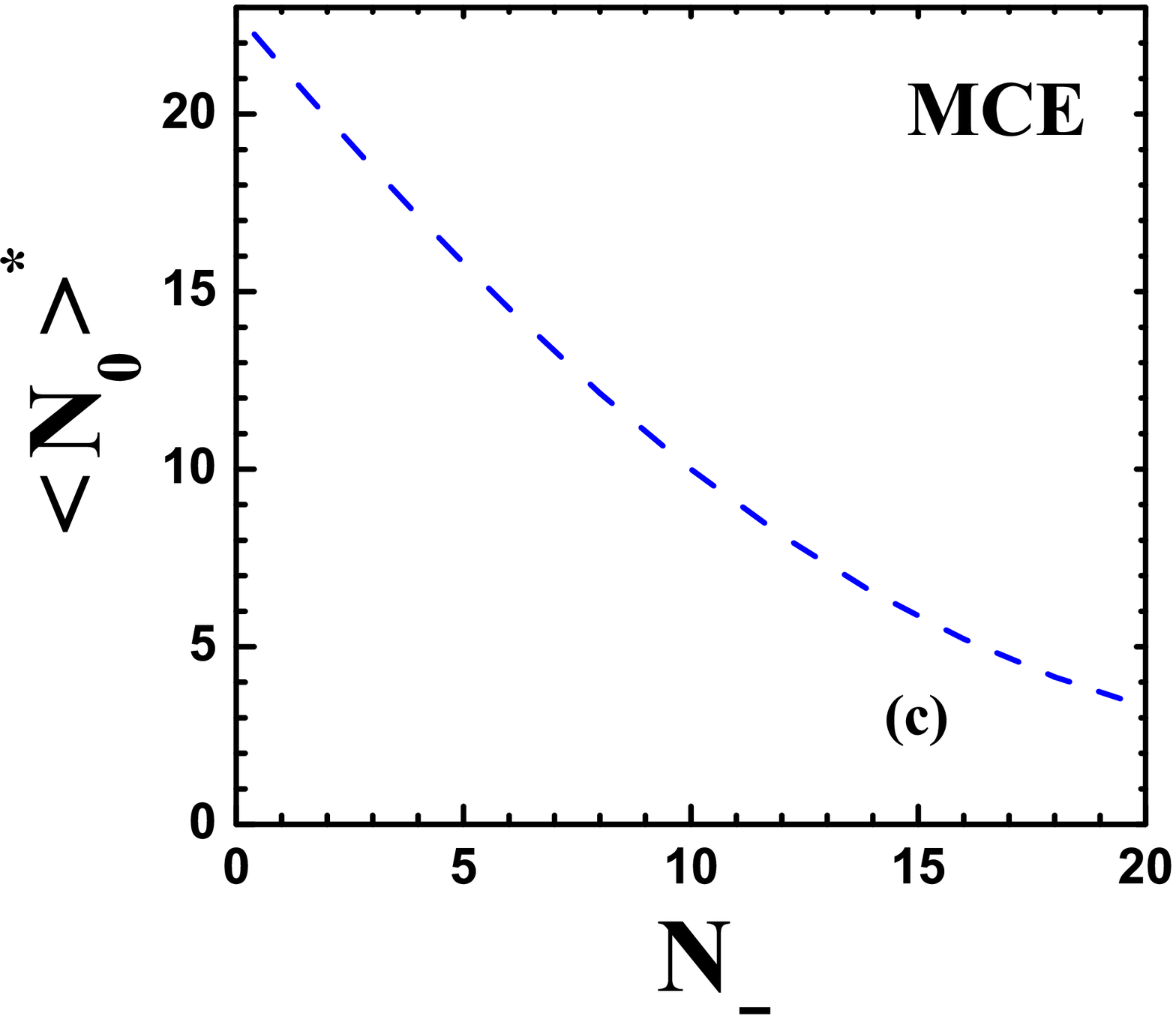,width=0.48\textwidth}
 \epsfig{file=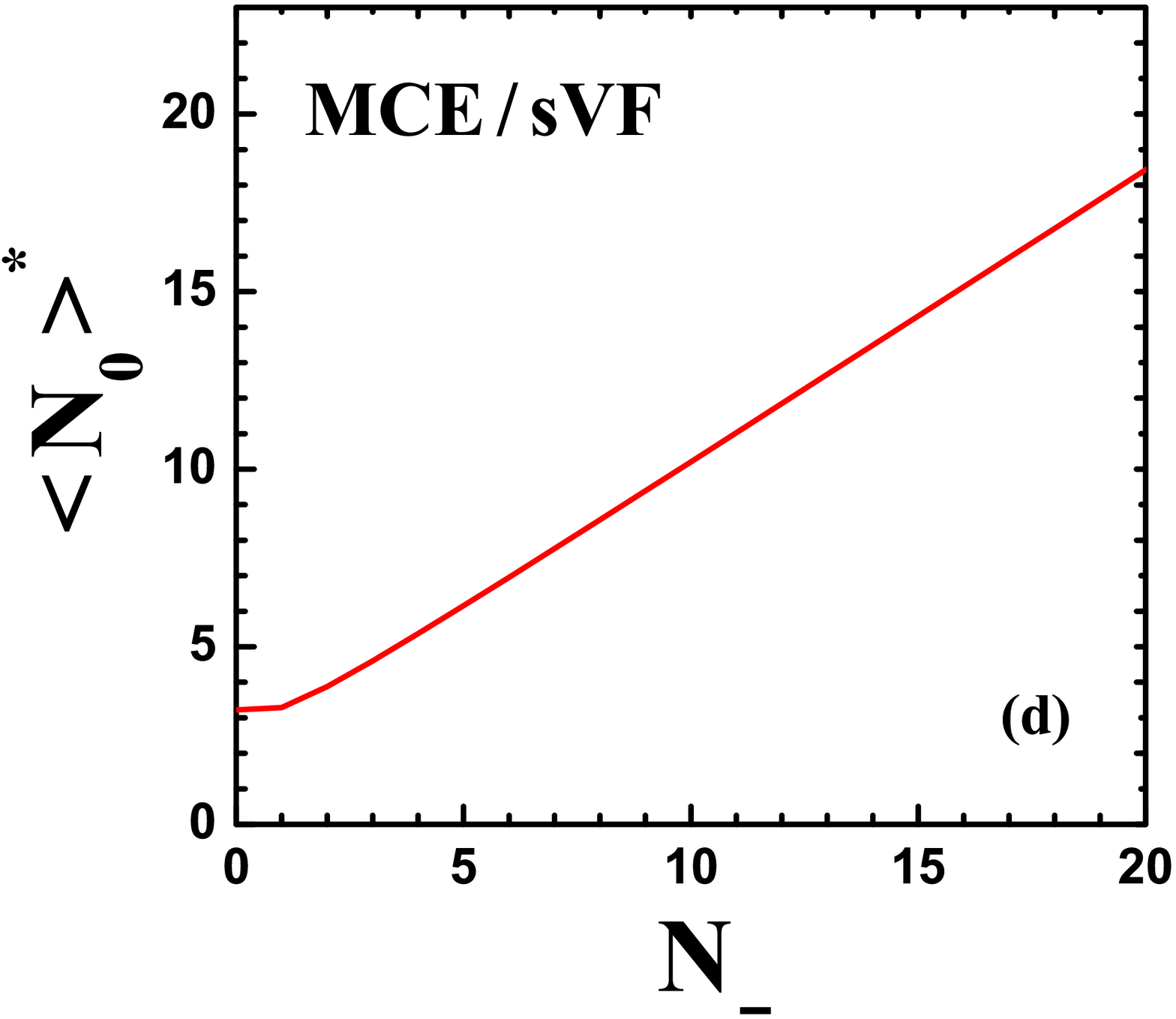,width=0.48\textwidth}
 \end{center}
 \vspace{-0.7cm}
 \caption{(Color online)
Examples of the dependence of the neutral particle mean multiplicity
on the multiplicity of negatively charged particles calculated within the
GCE ({\it top left}), CE ({\it top right}), MCE ({\it bottom left})
and MCE/sVF ({\it bottom right}).
The distributions are calculated assuming $\overline{N}=10$ (see text for details).
 \label{fig-N0}}
\end{figure}

The distributions in Fig.~\ref{fig-P} are rather different. Thus,
it is obvious that the dependence on $N_-$ of
the semi-inclusive mean multiplicity of neutral
particles defined as:
 \eq{\label{N0}
 \langle N_0\rangle^*
 \;\equiv\; \frac{\sum_{N_0} N_0~ P(N_0,N_-)}
       {\sum_{N_0}P(N_0,N_-)}~,
 }
is different in various ensembles.
Namely, it is independent of $N_-$ in the GCE and CE:
 \eq{\label{N0gce}
 \langle N_0\rangle^*_{gce}~=~\langle N_0\rangle^*_{ce}~=~\overline{N}~.
 }
In the MCE, $\langle N_0\rangle^*_{mce}$  monotonically decreases
with increasing $N_-$ and equals approximately (see Appendix
\ref{app-C}):
 \eq{\label{N0mce}
 \langle N_0\rangle^*_{mce}
 \;\cong\; \overline{N}\left(\frac{4}{3}~-~\frac{N_-}{3\overline{N}}\right)^3~.
 }
Finally, the positive correlation between $N_0$ and $N_-$ in the
MCE/sVF leads to an approximately linear increase\footnote{A
linear increase of $\langle N_0\rangle^*_{mce}$  with $N_-$ is due to
the assumption of massless particles. For non-zero value of mass $m$ the
relation (\ref{N0a}) is changed at large $N_-$. The maximum value of $N_-$ is
$N_-^{max}=E/2m$.  Eq.~(\ref{N0a}) remains
approximately valid for  $N_-\ll N_-^{max}$, but $\langle
N_0\rangle^*_{\alpha}$ approaches zero at $N_-\rightarrow
N_-^{max}$.} $\langle N_0\rangle^*_{\alpha}$ with increasing
$N_-$:
\eq{\label{N0a}
 \langle N_0 \rangle ^* _{\alpha}
 \;\cong\; N_-~.
}
The semi-inclusive mean multiplicities of neutral particles
calculated within  the GCE, CE (\ref{N0gce}), MCE (\ref{N0mce}),
and MCE/sVF (\ref{N0a}) are shown as  functions of $N_-$ in
Fig.~\ref{fig-N0}.

\subsection{Quantum Statistics}
In this Subsection we illustrate the effects of quantum statistics
in the GCE, CE, and MCE  using the micro-correlator method of
Ref.~\cite{steph}.
%The modified version of this method allows to
%calculate the fluctuations and correlations in the system with
%imposed exact conservation laws in thermodynamic limit,
%$V\rightarrow \infty$, see  \cite{CE2}. It was shown in \cite{CE}
%and \cite{mce1} that charge and energy conservation almost do not
%affect mean multiplicity if it is high enough, e.g. $\langle
%N\rangle\sim 10$. Thus,
The mean multiplicities in the CE or MCE are approximately the
same as in the GCE. They are given by the sum of mean occupation
numbers with momentum ${\bf p}$ \cite{landau}:
 \eq{ \label{np-aver-ch}
 \langle N^a\rangle_{ce} \;\cong\; \langle N^a\rangle_{mce}
 \;\cong\; \langle N^a\rangle_{gce}
 \;\equiv\; \sum_{\bf p}\langle n_{\bf p}^a \rangle
 ~=~ \sum_{\bf p}\frac{1}{\exp\left(\epsilon_{\bf p}/T\right)~-~
 \gamma}~.
 }
where $a$ is $+$, $-$, or $0$ and denotes positive, negative or
neutral particles, $\epsilon_{\bf p}=p$ is one particle energy for
massless particles, $\gamma=+1$ for Bose statistics, $\gamma=-1$
for Fermi statistics, and $\gamma=0$ corresponds to the Boltzmann
approximation used throughout  the present paper. We study a
neutral system, thus, chemical potentials are zero in
Eq.~(\ref{np-aver-ch}), and the average occupation numbers
$\langle n_{\bf p}^a \rangle\equiv \langle n_{\bf p}\rangle$ are
therefore the same for neutral and charged particles. The
summation over discrete levels can be substituted by the
integration in the thermodynamic limit:
%$V\rightarrow\infty$:
%
\eq{\sum_{\bf
p}~...~\cong~\frac{V}{2\pi^{2}}\int_{0}^{\infty}p^{2}dp~...~.}
The fluctuations and correlations in the GCE, CE and MCE are very
different, nevertheless they can be expressed in terms of the
fluctuations of the occupation numbers of a single momentum level
in the GCE,
 \eq{ \label{np-fluc-ch}
 \langle (\Delta n_{\bf p}^a)^2 \rangle_{gce}~\equiv ~ v_{\bf p}^{a\,2}~
%\equiv~\langle \left(n_{\bf p}^a\right)^{2}\rangle_{gce}~-~
%\langle n_{\bf p}^a\rangle_{gce}^{2}~
=~ \langle n_{\bf p}\rangle \left( 1~ +~\gamma \langle n_{\bf p}
\rangle \right)~
 . }
This is a main advantage of the micro-correlator method. It allows
to calculate the fluctuations and correlations using the following
micro-correlators  \cite{CE2}:
\eq{\label{corr-gce}
   \langle \Delta n_{\bf p}^a \Delta n_{\bf k}^{b} \rangle_{gce}
  &~=~ v^{a\,2}_{\bf p} \delta_{\bf pk}\,\delta_{ab}~,
  \\
   \label{corr-ce}
   \langle \Delta n_{\bf p}^a \Delta n_{\bf k}^{b} \rangle_{ce}
  &~=~ v^{a\,2}_p \delta_{\bf pk}\,\delta_{ab}
   ~-~  q^a q^b\,\frac{v^{a\,2}_{\bf p} v^{b\,2}_k}
        {\sum_{{\bf p},a} v^{a\,2}_{\bf p} q^{\alpha 2}}~,
   \\
   \label{corr-mce}
   \langle \Delta n_{\bf p}^a \Delta n_{\bf k}^{b} \rangle_{mce}
  &~=~ v^{a\,2}_{\bf p} \delta_{\bf pk}\,\delta_{ab}
   ~-~ \frac{v^{a\,2}_{\bf p} v^{b\,2}_{\bf k}}{|A|}
   \left[
   q^a q^b\sum_{{\bf p},a}v^{a\,2}_{\bf p} \epsilon_{\bf p}^2 ~ + ~
   \epsilon_{\bf p} \epsilon_{\bf k} \sum_{{\bf p},a} v^{a\,2}_{\bf p} q^{\alpha 2}
   \right]~,
   }
where $\delta_{\bf pk}$ and $\delta_{ab}$ are the Kronecker delta
symbols, $q^a$, $ q^b$ are particle charges, $\pm 1$ or $0$, and
   \eq{
   |A|~ \equiv ~ \left(\sum_{p,a}v^{a\,2}_{\bf p} \epsilon_{\bf p}^2\right)
   \cdot \left(\sum_{p,a}
   v^{a\,2}_{\bf p}  q^{\alpha 2}\right)~
   }
is the correlation determinant.
% that appears if one considers more
%then one conserved quantity, i.e. the exact conservation of
%several charges in the CE or energy and charge conservation in the
%MCE.

The variance and correlations in the GCE, CE and MCE are
calculated as the sums (integrals) over momentum of the
corresponding micro-correlators (\ref{corr-gce}-\ref{corr-mce}):
\eq{
    \langle(\Delta N_a^{2}) \rangle
    \;=\; \sum_{{\bf p},{\bf k}} \langle \Delta n_{\bf p}^a \Delta n_{\bf k}^a\rangle\;,
    \qquad
    \langle \Delta N_a\,\Delta N_b \rangle
    \;=\; \sum_{{\bf p},{\bf k}} \langle \Delta n_{\bf p}^a \Delta n_{\bf k}^{b}\rangle\;.
 }
%
%The influence of Bose and Fermi statistic on mean multiplicity in
%neutral system is not large: $\langle N^{Bose}\rangle/\langle
%N^{Boltz}\rangle\sim 1.2$ and $\langle N^{Fermi}\rangle/\langle
%N^{Boltz}\rangle\sim 0.9$ \cite{CE3}. Quantum statistic also
%slightly change scaled variance\footnote{The effects of quantum
%statistic are much stronger for fluctuations in charged system.
%The scaled variance may rise up to infinity near the point of
%Bose-condensation for $m\rightarrow \mu$ in the GCE \cite{CE3}.}.
%Using the micro-correlator method for the considered system of
%neutral and charged particles
One obtains for Bosons:
 \eq{\label{w-gce-Bose}
 \omega_{gce}^{\pm\,Bose}
 & \;=\; \frac{\sum_{\bf p} v_{\bf p}^2}{\sum_{\bf p} \langle n_{\bf p}\rangle}
   \;\cong \; \frac{\int_0^{\infty}p^2dp\;e^{p/T}\left(e^{p/T}-1\right)^{-2}}
                  {\int_0^{\infty}p^2dp\left(e^{p/T}-1\right)^{-1}}
   \;=\; \frac{\pi^2}{6\,\zeta(3)}
   \;\cong\; 1.368\;,
 \\
 \label{w-ce-Bose}
 \omega_{ce}^{\pm\,Bose}
 & \;=\; \frac{\sum_{\bf p} v_{\bf p}^2}{\sum_{\bf p} \langle n_{\bf p}\rangle}
   \;-\; \frac{\left(\sum_{\bf p}v^2_{\bf p}\right)^2}
         {\sum_{\bf p} \langle n_{\bf p}\rangle\sum_{p,a} v^{2}_{\bf p}\, q^{\alpha 2}}
   \;=\; \frac{\sum_{\bf p} v_{\bf p}^2}{2\sum_{\bf p} \langle n_{\bf p}\rangle}
   \;=\; \frac{1}{2}\,\omega_{gce}^{\pm\,Bose}
  \;\cong\; 0.684\;,
 \\
 \label{w-mce-Bose1}
 \omega_{mce}^{\pm\,Bose}
 & \;=\; \frac{\sum_{\bf p} v_{\bf p}^2}{\sum_{\bf p} \langle n_{\bf p}\rangle}
   \;-\; \frac{\left(\sum_{\bf p}v^2_{\bf p}\right)^2}
         {\sum_{\bf p} \langle n_{\bf p}\rangle\sum_{p,a} v^{2}_{\bf p}\, q^{\alpha 2}}
   \;-\; \frac{\left(\sum_{\bf p}v^2_{\bf p}\,\epsilon_{\bf p}\right)^2}
         {\sum_{\bf p} \langle n_{\bf p}\rangle\sum_{\bf{p},a} v^{2}_{\bf p}\,\epsilon_{\bf p}^2}
 %  \\
 %& \;=\; \frac{\sum_{\bf p} v_{\bf p}^2}{2\sum_{\bf p} \langle n_{\bf p}\rangle}
 %  \;-\; \frac{\left(\sum_{\bf p}v^2_{\bf p}\,\epsilon_{\bf p}\right)^2}
 %        {3\sum_{\bf p} \langle n_{\bf p}\rangle\sum_{\bf p} v^{2}_{\bf p}\,\epsilon_{\bf p}^2}
 %  \nonumber \\
 %  \label{w-mce-Bose}
 %& \;=\; \frac{1}{2}\,\omega_{gce}^{\pm\,Bose}
 %  \;-\; \frac{\left(\int_0^{\infty}p^3dp\;e^{p/T}\left(e^{p/T}-1\right)^{-2}\right)^2}
 %             {3\int_0^{\infty}p^2dp\left(e^{p/T}-1\right)^{-1}
 %               \int_0^{\infty}p^4dp\;e^{p/T}\left(e^{p/T}-1\right)^{-2}}
   \nonumber\\
 & \;=\; \frac{\pi^2}{12\,\zeta(3)} \;-\; \frac{45\,\zeta(3)}{2\,\pi^4}
   \;\cong \; 0.407\;,
 }
where $\zeta(3)\cong 1.202$ is the zeta Riemann function. In
calculating (\ref{w-gce-Bose}-\ref{w-mce-Bose1}) we use $\langle
n_{\bf p}^a \rangle\equiv \langle n_{\bf p}\rangle$ and $v_{\bf
p}^{+\,2}=v_{\bf p}^{-\,2} =v_{\bf p}^{0\,2} \equiv v_{\bf p}^2$
for a neutral system. This gives: $\sum_{{\bf p},a}v^{a\,2}_{\bf
p}\, q^{a\,2}=2\sum_{\bf p}v^2_{\bf p}$ and $\sum_{{\bf
p},a}v^{a\,2}_{\bf p}\,\epsilon_{\bf p}^2=3\sum_{\bf p}v^2_{\bf
p}\,\epsilon_{\bf p}^2$. Similarly one can get the results for
Fermions:
%, replacing $-1$ by $+1$ in
%Eqs.~(\ref{w-gce-Bose}-\ref{w-mce-Bose1}):
%
 \eq{
 \omega_{gce}^{\pm\,Fermi} &\;=\; \frac{\pi^2}{9\,\zeta(3)} \;\cong \; 0.912\;,
 \\
  \omega_{ce}^{\pm\,Fermi} &\;=\; \frac{1}{2}\,\omega_{gce}^{\pm\,Fermi}
  \;\cong \; 0.456 \;,
 \\
  \omega_{mce}^{\pm\,Fermi} &\;=\; \frac{\pi^2}{18\,\zeta(3)}
  \;-\; \frac{135\,\zeta(3)}{7\,\pi^4}
  \;\cong \; 0.218\;.
 }

The scaled variances for neutral particles are:
% The
%only difference is that
%charge conservation do not affect the fluctuations of neutral
%particles. It leads to the absence of the second therm in the
%Eqs.~(\ref{w-ce-Bose}), (\ref{w-mce-Bose1}) and to the following
%results:
%
 \eq{
 \omega^{0\,Bose}_{gce}
 &\;=\; \omega^{0\,Bose}_{ce} \;=\; \omega^{\pm\,Bose}_{gce} \;\cong \; 1.368\;,
 %\\
~~~~
 \omega^{0\,Fermi}_{gce}
 \;=\; \omega^{0\,Fermi}_{ce} \;=\; \omega^{\pm\,Fermi}_{gce} \;\cong\; 0.912\;,
 \\
 \omega^{0\,Bose}_{mce}
 &\;= \; \frac{\pi^2}{6\,\zeta(3)} \;-\; \frac{45\,\zeta(3)}{2\,\pi^4}
  \; \cong \; 1.091\;,
 %\\
 ~~~~
 \omega^{0\,Fermi}_{mce}
 \;= \; \frac{\pi^2}{9\,\zeta(3)} \;-\; \frac{135\,\zeta(3)}{7\,\pi^4}\;
 \cong\;
 0.674\; .
 }

The correlation coefficients can be also calculated using
micro-correlator method:
 \eq{
 \rho^{a\,b} \;=\; \frac{\langle \Delta N_a\,\Delta N_b \rangle}
 {\sqrt{\omega^a\cdot\langle N_a\rangle\cdot\omega^b\cdot\langle N_b \rangle}}
 \;=\; \frac{1}{\sqrt{\omega^a\,\omega^b}\;}
       \frac{\sum_{{\bf p},{\bf k}} \langle \Delta n_{\bf p}^a \Delta n_{\bf k}^{b}\rangle}
           {\sum_{\bf p}\langle n_{\bf p}\rangle}\;.
 }
%
%Similarly to the calculation of scaled variances one can find that
There are no correlations between positively and negatively
charged particles in the GCE, $\rho^{+-}_{gce}=0$, and there is
the absolute correlation in the CE and MCE,
$\rho^{+-}_{ce}=\rho^{+-}_{mce}=1$. These values are the same for
any type of statistics. The correlation between charged and
neutral particles, $\rho^{0-}=\rho^{0+}$ (\ref{Rho}),
%is also different. It
is zero for the GCE and CE, but it has a negative value for the
MCE.
% This means an anti-correlation because of the exact energy
%conservation.
The correlation coefficient $\rho^{0-}_{mce}$ reads:
 \eq{\label{rho-mce-om}
 \rho_{mce}^{0-}
 &\;=\; -\;\frac{1}{\sqrt{\omega^0\,\omega^-}}~~
       \frac{\left(\sum_{\bf p}v^2_{\bf p}\,\epsilon_{\bf p}\right)^2}
         {3\sum_{\bf p} \langle n_{\bf p}\rangle\sum_{\bf p} v^{2}_{\bf p}\,\epsilon_{\bf
         p}^2}~.
         }
The Eq.~(\ref{rho-mce-om}) gives for Bosons and Fermions:
\eq{
 \rho_{mce}^{0-\,Bose}
 &\;=\; \;-\; \sqrt{2}\left[
              \left(\frac{\pi^6}{135\,\zeta(3)^2} - 2\right)
              \left(\frac{\pi^6}{135\,\zeta(3)^2} - 1\right)\right]^{-1/2}
 \;\cong \; -~0.417 \;,
        \\
 \rho_{mce}^{0-\,Fermi}
 &\;=\; \;-\; \sqrt{2}\left[
              \left(\frac{7\pi^6}{1215\,\zeta(3)^2} - 2\right)
              \left(\frac{7\pi^6}{1215\,\zeta(3)^2} - 1\right)\right]^{-1/2}
 \;\cong \; -~0.621 \;.\qquad
}

The scaled variances and correlation coeffients for the Boltzmann
approximation can be obtained from
Eqs.~(\ref{w-gce-Bose}-\ref{w-mce-Bose1},\ref{rho-mce-om})
replacing $\gamma=0$ in Eq.~(\ref{np-aver-ch}):
\eq{
\omega_{gce}^{\pm\,Boltz}~&=~1~,~~~~
\omega_{ce}^{\pm\,Boltz}~=~\frac{1}{2}~,~~~~\omega_{mce}^{\pm\,Boltz}~=~\frac{1}{4}~,
\\
%~~~~
 \omega^{0\,Boltz}_{gce}~&=~\omega^{0\,Boltz}_{ce}~=~1~,~~~~
\omega^{0\,Boltz}_{mce}~=~\frac{3}{4}~,~~~~
%\\
 \rho_{mce}^{0-\,Boltz} \;=\; -~\frac{1}{\sqrt{3}}
 \;\cong \; -~0.577 \;.
 }
%
%where $\rho_{mce}^{0-\,Boltz}$ coincide with our previous result
%for Boltzmann statistic .
They, of course,  coincide with our previous results
(\ref{w-gce-Boltz}-\ref{w-ce-Boltz},\ref{w-mce-Boltz}-\ref{Rho0m})
for the Boltzmann statistics.
% see Eqs.~. instead of , see Eqs.~(\ref{w-gce-Boltz}),
%(\ref{w-ce-Boltz}), (\ref{w-mce-Boltz}).

One can conclude that Bose statistics makes the fluctuations
always bigger and Fermi statistics -- smaller:
$\omega^{Fermi}<\omega^{Boltz}<\omega^{Bose}$, in all statistical
ensembles (GCE, CE, MCE) and for all types of particles (positive,
negative, neutral). The strongest effect for the neutral system in
equilibrium \footnote{The effects of quantum statistics for
fluctuations can be much stronger at non-zero chemical potential.
The scaled variance of Bosons may rise up to infinity near the
point of Bose-condensation \cite{CE3}.} is for the scaled variance
of charged Bosons in the MCE:
$\omega_{mce}^{\pm\,Bose}/\omega_{mce}^{\pm\,Boltz}\cong 1.6$. The
only correlation coefficient which fills an influence of quantum
statistics is $\rho^{0\pm}$ in the MCE:
$\rho_{mce}^{0\pm~Fermi}<\rho_{mce}^{0\pm~Boltz}<\rho_{mce}^{0\pm~Bose}<0$.
%Contrary, the anti-correlation for fermions is bigger than for
%bosons.
However, the quantum statistics does not change a sign of this
correlation.
%according to Eq.~(\ref{w-a}) these changes only
%slightly affect the MCE/sVF distribution.
Thus, the main features of the GCE, CE, and MCE fluctuations and
correlations found in the Boltzmann approximation -- constant
values of $\omega^{\pm}$ and $\omega^0$ in the thermodynamic
limit, strong  correlations, $\rho_{ce}^{+-}=\rho_{mce}^{+-}=1$,
due to the exact charge conservation,  and anti-correlation
between neutral and charged particles, $\rho_{mce}^{0\pm}<0$, due
to the exact energy conservation  -- remain the same for Bose and
Fermi statistics.
The quantum statistics can not simulate the MCE/sVF effects: an
increase of the scaled variances  in proportion to the mean
multiplicities (\ref{w-a}) and a strong {\it positive}
correlation, $\rho_{\alpha}^{0\pm}\cong 1,$ between neutral and
charged particles (\ref{N0a}). These new effects take place due to
the scaling volume fluctuations in the MCE/sVF.

\section{semi-inclusive  momentum spectra}

In this section  single particle momentum spectra of negatively
charged particles are considered. The inclusive spectra are
denoted as $F(p)$ and the semi-inclusive at fixed $N_-$ as
$F^*(p)$.  In both cases the spectra are normalized to unity:
$\int_0^{\infty}p^2dp\; F(p)=1$ and $\int_0^{\infty}p^2dp\;
F^{*}(p)=1$.

\subsection{GCE and CE}
The inclusive and semi-inclusive momentum spectra  in the GCE and
CE are equal and read\footnote{This is true for the used here
Boltzmann statistics. The form of momentum spectrum in the CE
becomes different from that in the GCE for quantum gases in
finite volumes. For the isospin conservation this was demonstrated
in Ref.~\cite{MR}}
 \eq{\label{Fp}
F_{gce}(p)
 ~=~ F^*_{gce}(p)~=~F_{ce}(p)~=~ F^*_{ce}(p)~=~
  \frac{1}{2T^3}~\exp\left(-~\frac{p}{T}\right)\;.
 }
This follows from the fact that a single particle momentum
spectrum and the particle multiplicity are uncorrelated in these
ensembles.

\subsection{Micro-Canonical Ensemble}
The inclusive single particle momentum spectrum of negatively
charged particles in the MCE reads:
 \eq{\label{Fp-mce}
 F_{mce}(p)
\;=\; \frac{1}{\overline{N}}~\frac{1}{2E^3}
        \sum_{N_0=0}^{\infty}\sum_{N_-=1}^{\infty} \frac{N_-~(3N_0+6N_--1)!}
        {(3N_0+6N_--4)!}\,\left(1~-~\frac{p}{E}\right)^{3N_0+6N_--4}
        P_{mce}(N_0,N_-)\;,
 }
see also Ref.~\cite{powerlaw2}.
\begin{figure}[ht!]
 \begin{center}
 \epsfig{file=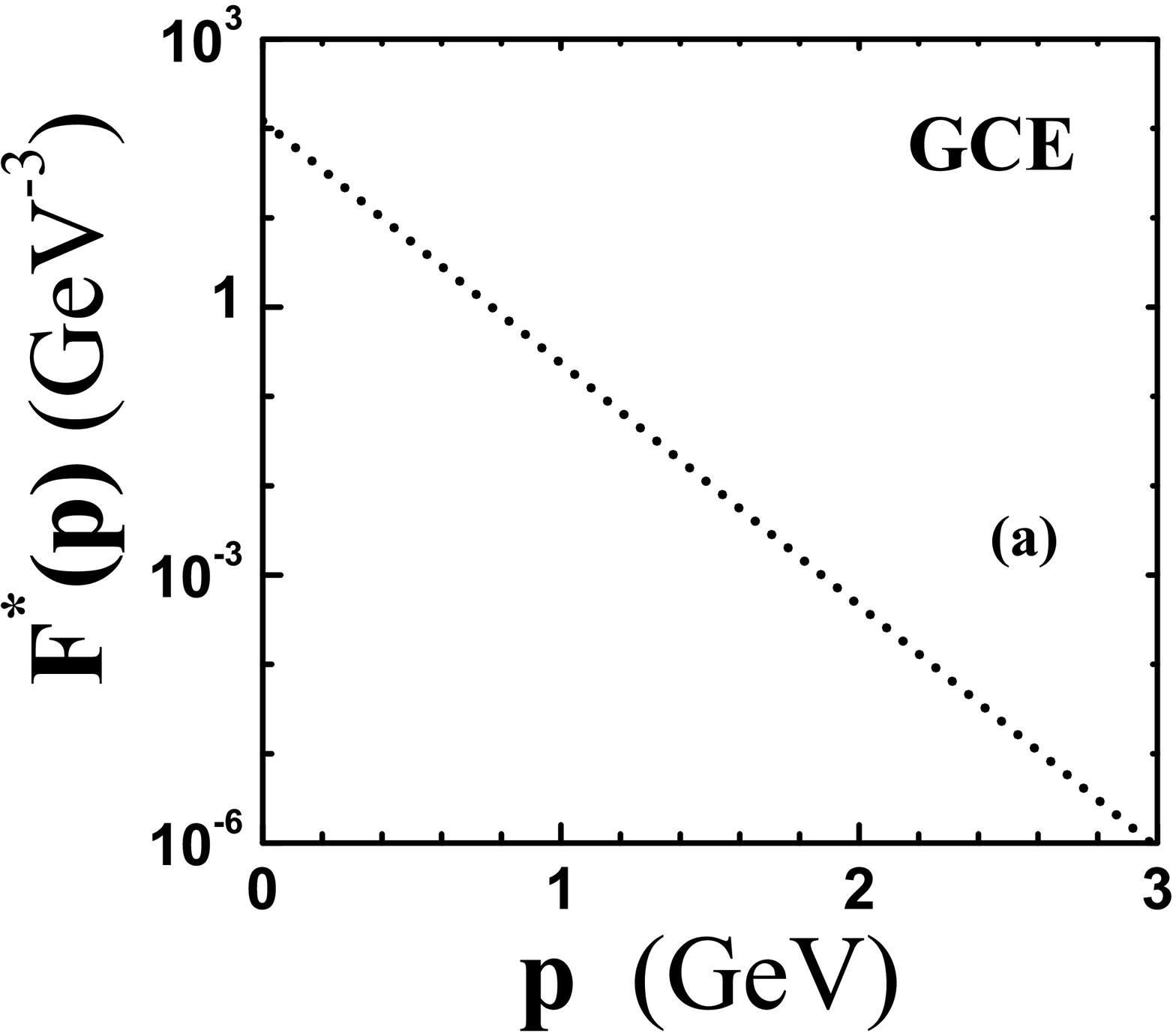,width=0.48\textwidth}\;\;
\epsfig{file=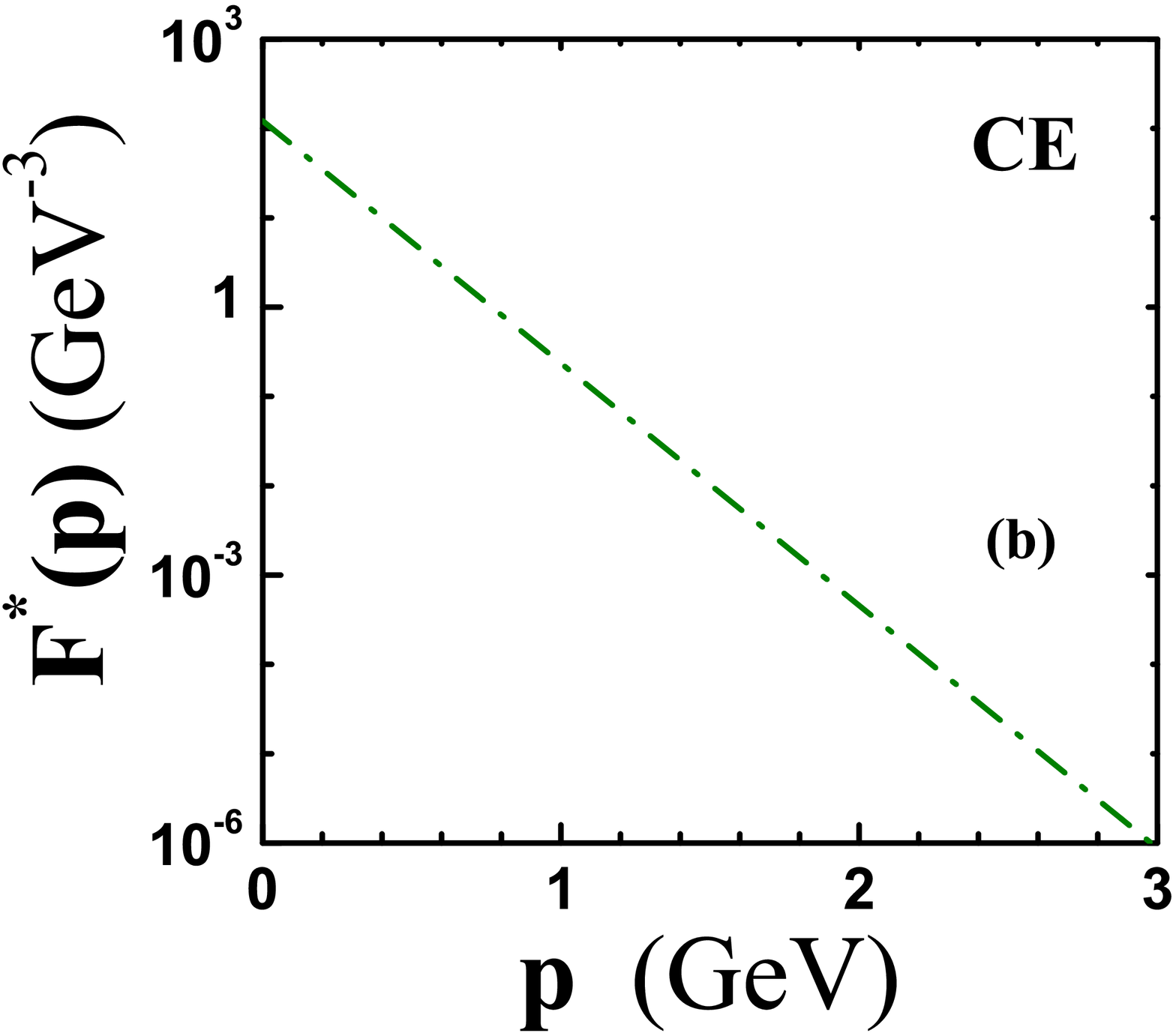,width=0.48\textwidth}\;\;
 \epsfig{file=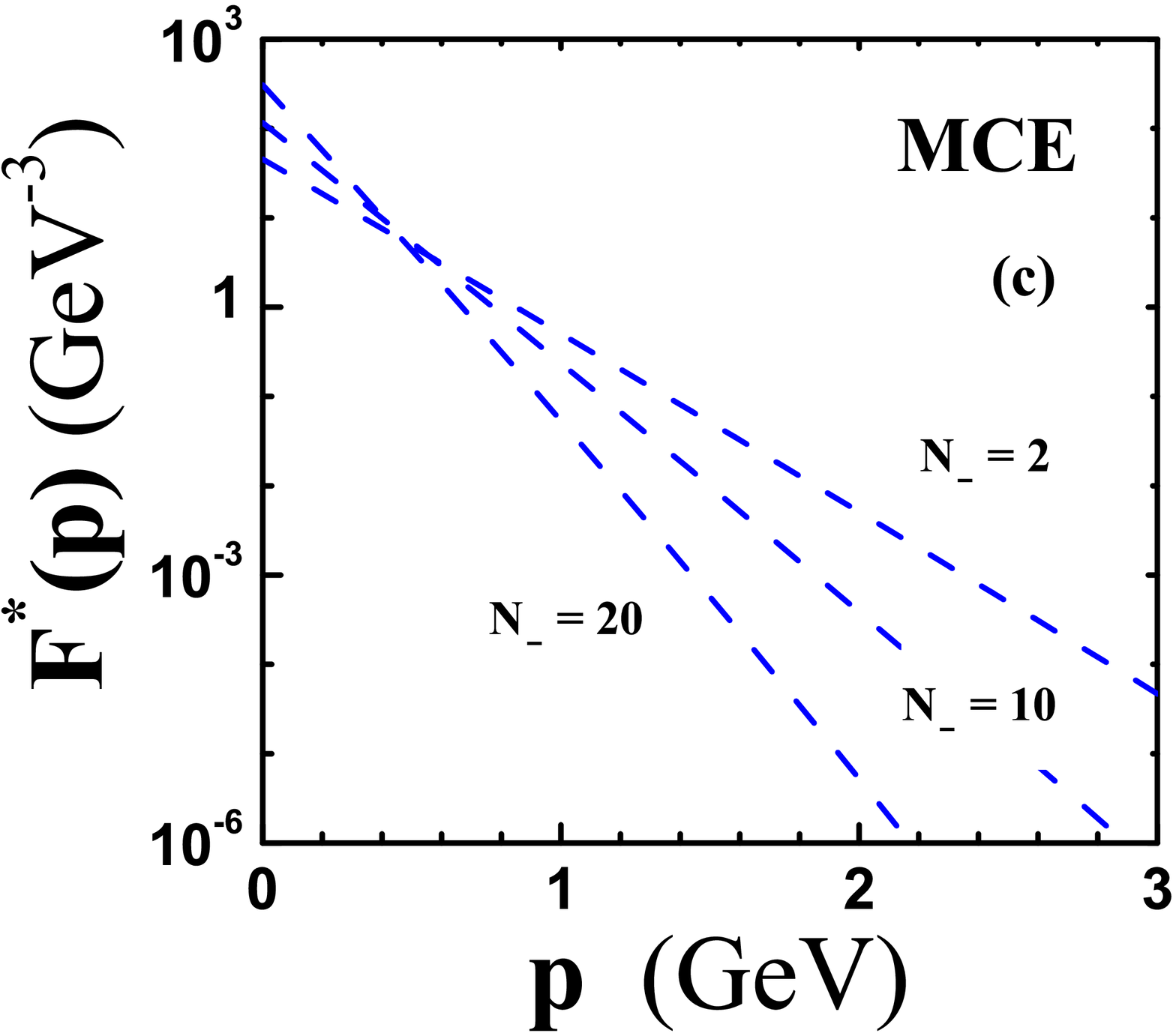,width=0.48\textwidth}
 \epsfig{file=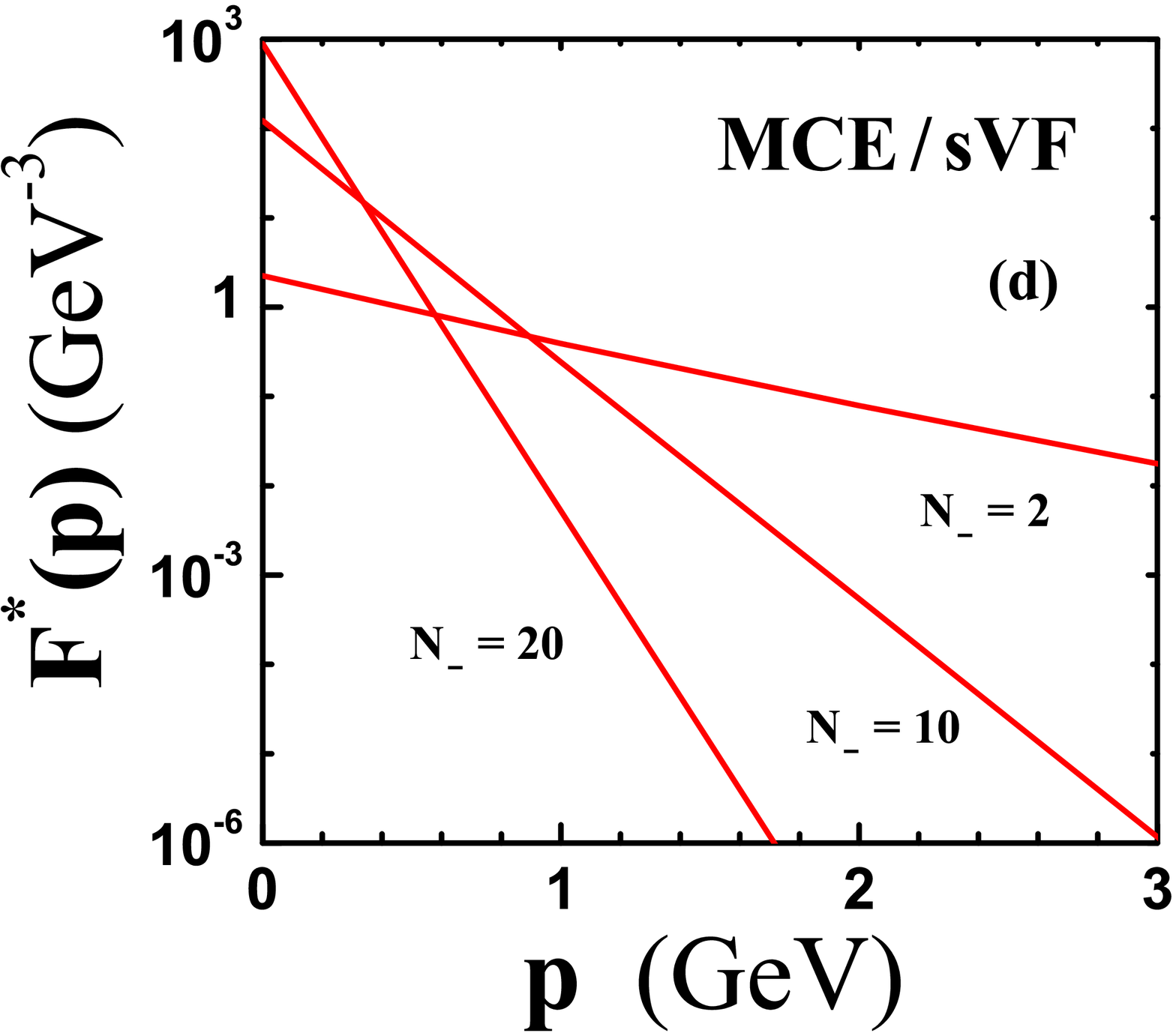,width=0.48\textwidth}\;\;
 \end{center}
 \vspace{-0.7cm}
 \caption{(Color online)
Examples of the semi-inclusive momentum spectra  of negatively
charged particles calculated within the GCE ({\it top left}), CE
({\it top right}), MCE ({\it bottom left}) and MCE/sVF ({\it
bottom right}) for three values of $N_-$. The distributions are
calculated assuming $\overline{N}=10$ and $T=160$~MeV (see text
for details).
 \label{fig-Fp}}
\end{figure}
The spectrum (\ref{Fp-mce}) approximately has the Boltzmann form
(\ref{Fp}) at momenta $p$ significantly  smaller than the total system
energy $E$. However, large deviations from (\ref{Fp}) are observed
close to the threshold, $p=E$, where the MCE spectrum approaches
zero. The inclusive spectra $F(p)$ in the GCE, CE (\ref{Fp}), and
MCE (\ref{Fp-mce}) are shown in Fig.~\ref{fig-in}, {\it right}.
%same as in Ref.~\cite{powerlaw2}.

\begin{figure}[ht!]
 \begin{center}
 \epsfig{file=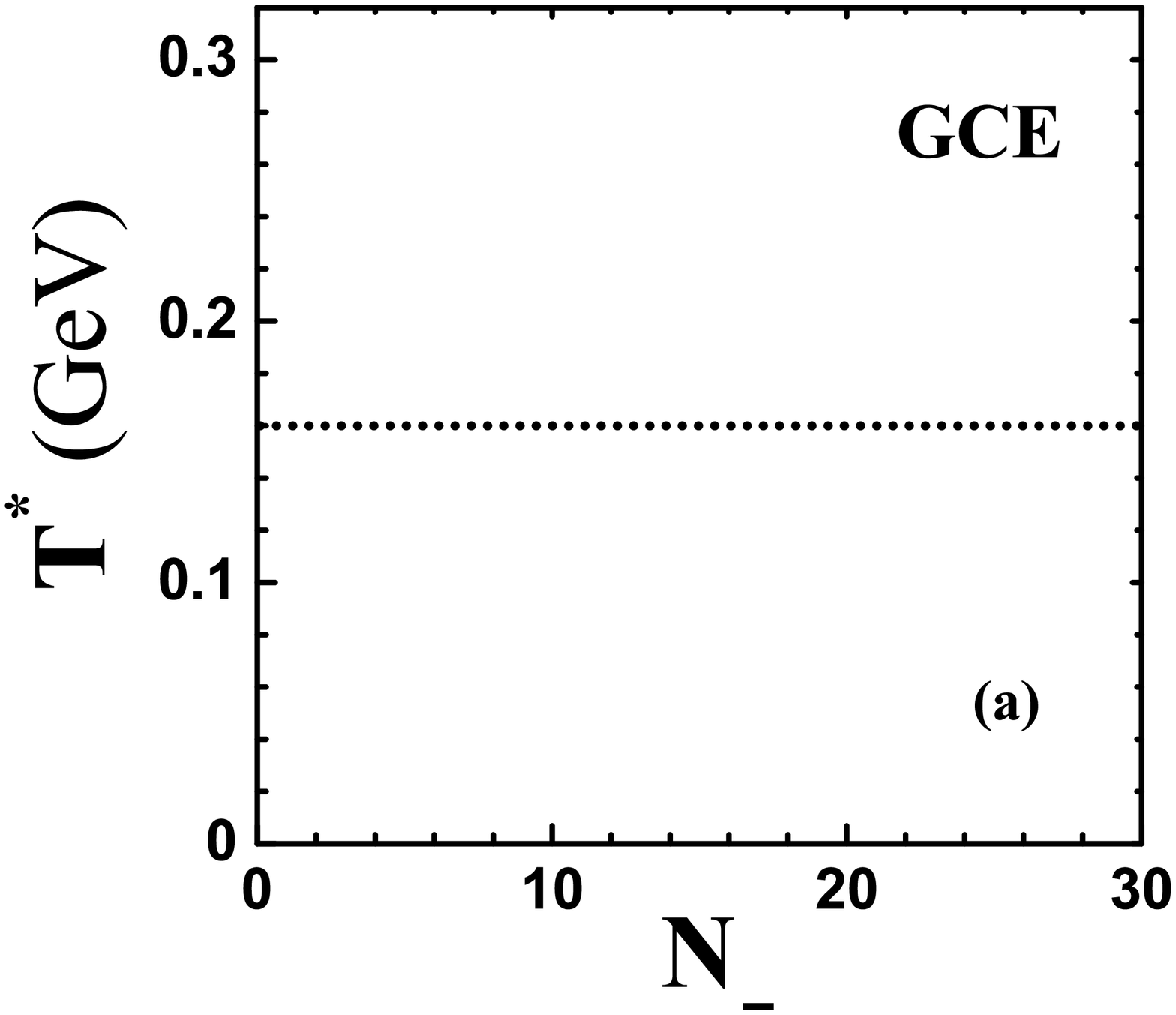,width=0.48\textwidth}\hspace{0.2cm}
 \epsfig{file=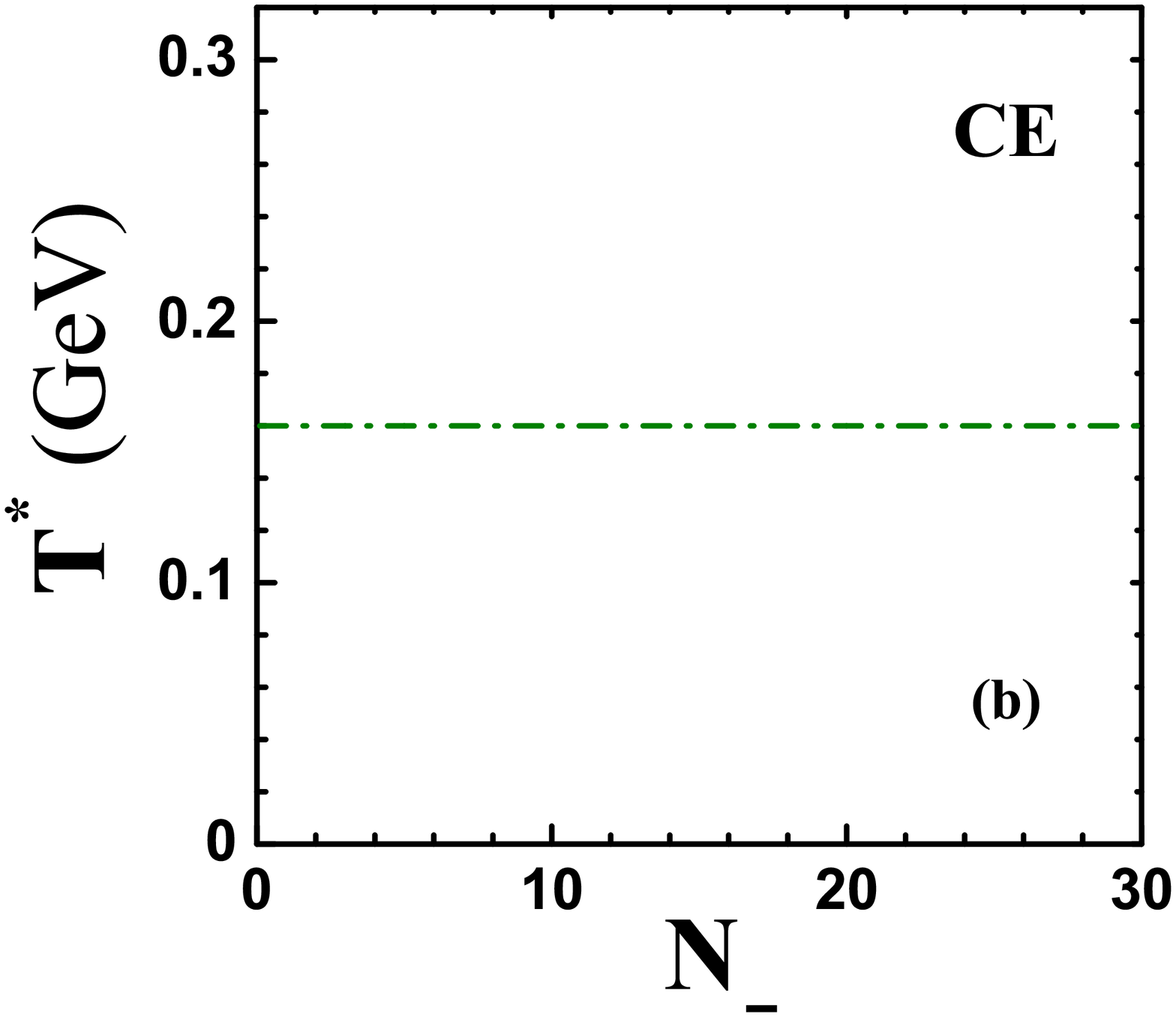,width=0.48\textwidth}\\[0.2cm]
 \epsfig{file=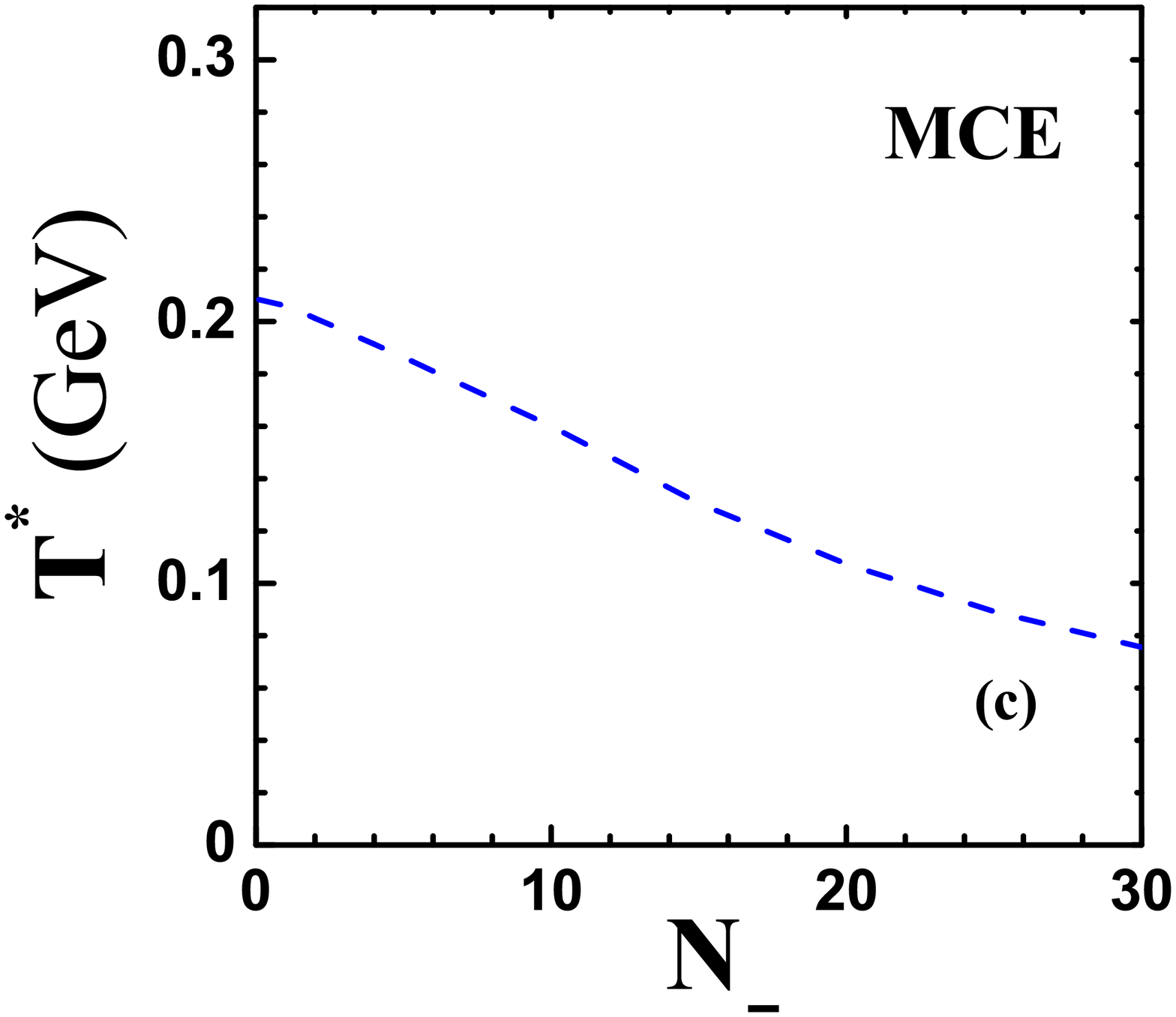,width=0.48\textwidth}\hspace{0.2cm}
 \epsfig{file=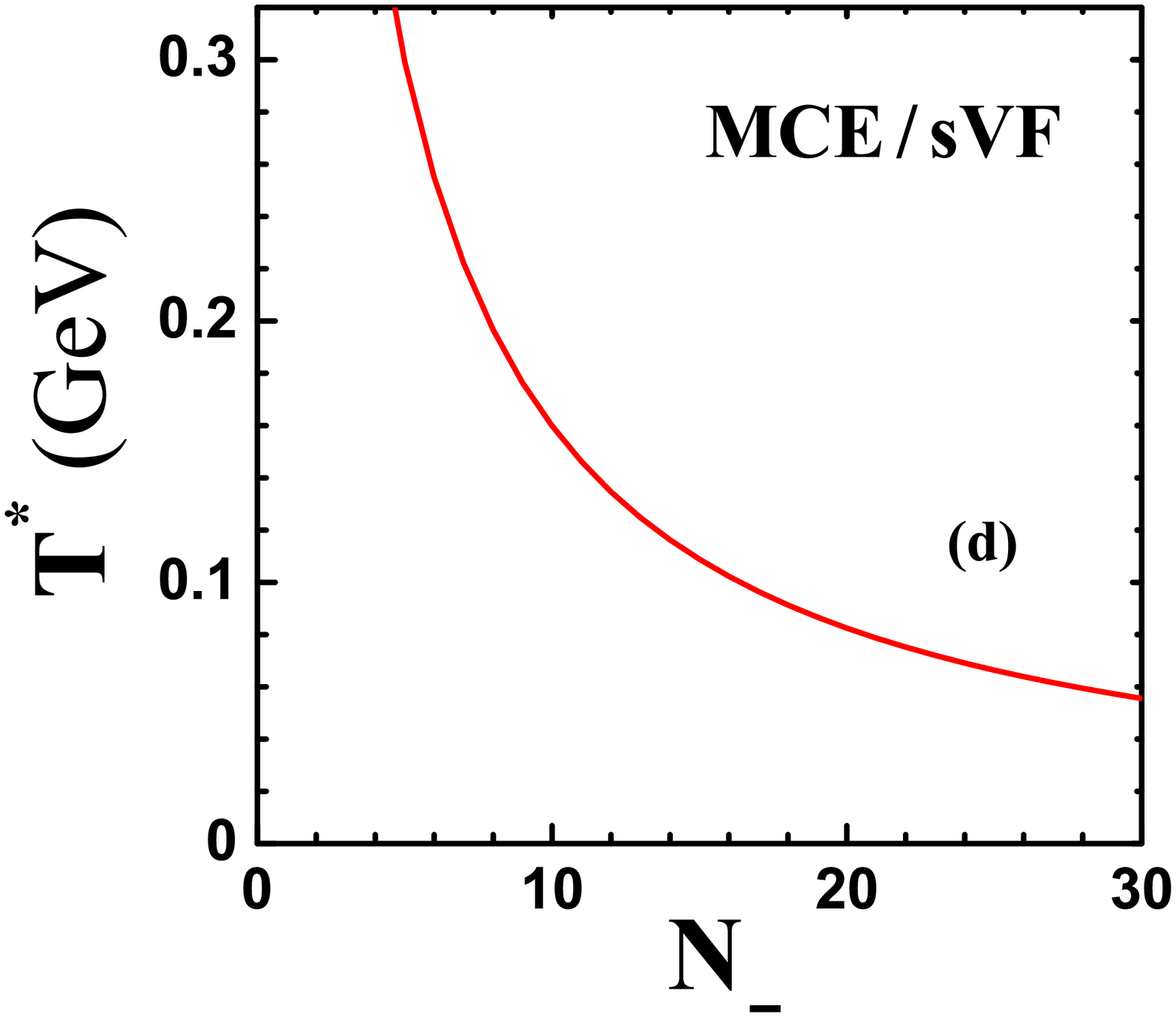,width=0.48\textwidth}
 \end{center}
 \vspace{-0.7cm}
 \caption{(Color online)
The dependence of the inverse slope parameter of
the momentum spectra on the multiplicity  of
negatively charged particles $N_-$ calculated within the
GCE ({\it top left}), CE ({\it top right}), MCE ({\it bottom left})
and MCE/sVF ({\it bottom right}).
The distributions are calculated assuming  $\overline{N}=10$ and $T=160$~MeV  (see text for details).
\label{fig-T*}}
\end{figure}

The semi-inclusive momentum spectrum at a fixed number of negatively charged
particles is given by:
 \eq{\label{F*mce}
 F^*_{mce}(p)
 \;=\; \frac{C}{2E^3}
        \sum_{N_0=0}^{\infty} \frac{(3N_0+6N_--1)!}
        {(3N_0+6N_--4)!}~\left(1~-~\frac{p}{E}\right)^{3N_0+6N_--4}
        ~P_{mce}(N_0,N_-)~,
 }
where $N_-\geq 1$ and $C=
\left[\sum_{N_0}P_{mce}(N_0,N_-)\right]^{-1}$ is the normalization
factor. Examples of the $F^*_{mce}(p)$ spectrum for three values
of $N_-$ are shown in Fig.~\ref{fig-Fp} {\it bottom, left}. The
semi-inclusive spectra in the MCE (\ref{F*mce}) have the Boltzmann
form for $p\ll E$,
\eq{\label{F*mceT}
F^*_{mce}(p)~ \cong
~\frac{1}{2T_{mce}^{*3}}~\exp\left(-~\frac{p}{T_{mce}^*}\right)~,
}
and the inverse slope parameter $T_{mce}^*$ depends on ~$N_-$.
This dependence is presented in Fig.~\ref{fig-T*} for the MCE and
other ensembles studied here.
%
%One can see that
In the GCE and CE the $T^*$ is independent of $N_-$ and
equal to the inverse slope parameter of the inclusive spectrum,
 $T^*_{gce}=T^*_{ce}=T=160$~MeV.
In the MCE the inverse slope parameter
decreases
with increasing  $N_-$
and it crosses the line
$T=160$~MeV at $N_-=\overline{N}$. Thus, the inclusive
momentum spectrum $F_{mce}(p)$ (\ref{Fp-mce}) coincides with the
semi-inclusive one, $F^*_{mce}(p)$ (\ref{F*mce}), at the crossing point.

%%%%%%%%%%%%%%%%%%%%%%%%%%%%%%%%%%%%%%%%%%%%%%%%%%%%%%%%%%%%%%%%%%%%%%%%%%
%
\subsection{MCE with scaling Volume Fluctuations}
%
%%%%%%%%%%%%%%%%%%%%%%%%%%%%%%%%%%%%%%%%%%%%%%%%%%%%%%%%%%%%%%%%%%%%%%%%%%
%
%
The inclusive single particle momentum spectrum in the MCE/sVF equals to:
%can be obtained in the same way as in Ref.~\cite{powerlaw2}:
%
 \eq{\label{Fp-a}
 F_{\alpha}(p)
 \;=\; \frac{1}{\overline{N}}~\frac{1}{2E^3}
       \sum_{N_0=0}^{\infty}\sum_{N_-=1}^{\infty} \frac{N_-~(3N_0+6N_--1)!}
 {(3N_0+6N_--4)!}~\left(1~-~\frac{p}{E}\right)^{3N_0+6N_--4}
 ~P_{\alpha}(N_0,N_-)~.
 }
The structure of Eq.~(\ref{Fp-a}) is the same as the structure of
the corresponding Eq.~(\ref{Fp-mce})  for the MCE. The only
difference is in the form of the multiplicity distribution, namely
$P_{\alpha}(N_0,N_-)$ is used in Eq.~(\ref{Fp-a}) instead of
$P_{mce}(N_0,N_-)$ in Eq.~(\ref{Fp-mce}). The inclusive spectrum
$F_{\alpha}(p)$ is shown in Fig.~\ref{fig-in} {\it right}.
It can be well approximated by
the power law dependence:
 \eq{\label{power}
 F_{\alpha}(p) \;\cong\;
 \frac{k^{k}\Gamma(k+4)}{2\Gamma(k)}\;T^{k+1}\;\left(p~+~Tk\right)^{-k-4}
 \;\cong~11.27~\rm{GeV}^5~ (p~+~4T)^{-8}\;,
 }
where  $k=4$ is used in the last expression (see Ref.~\cite{powerlaw2}).

The semi-inclusive momentum spectrum in the MCE/sVF reads:
 \eq{\label{F*a}
 F_{\alpha}^{*}(p)
 \;=\; \frac{C}{2E^3}
        \sum_{N_0=0}^{\infty} \frac{(3N_0+6N_--1)!}
        {(3N_0+6N_--4)!}~\left(1~-~\frac{p}{E}\right)^{3N_0+6N_--4}
        ~P_{\alpha}(N_0,N_-)~,
 }
where $C=\left[\sum_{N_0} P_{\alpha}(N_0,N_-)\right]^{-1}$.
The  spectrum $F_{\alpha}^{*}(p)$ is plotted in Fig.~\ref{fig-Fp} for
several  values of $N_-$.
Similar to the MCE spectrum (\ref{F*mceT})
the MCE/sVF one
can be approximated
as:
\eq{\label{F*aT}
F^*_{\alpha}(p)~ \cong
~\frac{1}{2T_{\alpha}^{*3}}~\exp\left(-~\frac{p}{T_{\alpha}^*}\right)~,
}
with the inverse slope parameter $T_{\alpha}^*$. The dependence of
$T_{\alpha}^*$ on $N_-$ is shown in Fig.~\ref{fig-T*} {\it bottom
right}. The MCE/sVF temperature $T_{\alpha}^*$ decreases with
increasing $N_-$.
For $N_-=\overline{N}$ the inverse slope parameter $T^*$ is the
same in the MCE and MCE/sVF and equals to the parameter $T$ in the
GCE and CE. The analytical approximations of the dependence of $T^*$
on $N_-$ in the MCE and MCE/sVF are presented in Appendix
\ref{app-C}.
%
%%%%%%%%%%%%%%%%%%%%%%%%%%%%%%%%%%%%%%%%%%%%%%%%%%%%%%%%%%%%%%%%%%%%%%%
%
\section{Comparison with data}\label{S-V}
A quantitative comparison of the discussed statistical models
with the experimental data requires a
significant additional effort, which is far beyond the scope of this paper.
In particular one should introduce
proper degrees of freedom and all related conservation laws
as well as a longitudinal collective motion of matter.
Nevertheless, a qualitative comparison seems to be useful already,
and consequently it is presented in this section.

An excellent review of the experimental data on semi-inclusive
properties of $p+p$ interactions at high energies can be found in
Ref.~\cite{ag_semi}.
The volume scaling function
as well as the value of the temperature
parameter used in this work in quantitative calculations
were selected in order to approximately reproduce the
results on the inclusive distributions in $p+p$ interactions.
Consequently, a comparisons between these data and the model results
is justified.
Clearly, as the influence of global conservation laws is crucial
for the considered statistical approaches the data referring to
the semi-inclusive properties measured in the full phase-space are
of primary importance. Several features of these data are well
established~\cite{ag_semi}. Two of them are relevant for the
comparison with the models discussed here, namely:

(A) the mean multiplicity of produced $\pi^0$ mesons increases with
increasing multiplicity of negatively charged particles and

(B) the average transverse momentum of negatively charged
particles decreases with increasing multiplicity of these
particles.

The property (B) needs several comments. First, it is well
experimentally established~\cite{ag_semi} for charged hadron
multiplicity and mean transverse momentum measured in full
phase-space in p+p interactions at 6.6-400~GeV/c. Clearly, the
full phase-space results are relevant when effects related to the
global conservation laws are of interest. Second, the  mean
transverse momentum increases with increasing multiplicity when
mid-rapidity values are considered~\cite{extra}. An interpretation
of this dependence is, however, beyond the scope of this paper.

\begin{figure}[ht!]
 \begin{center}
 \epsfig{file=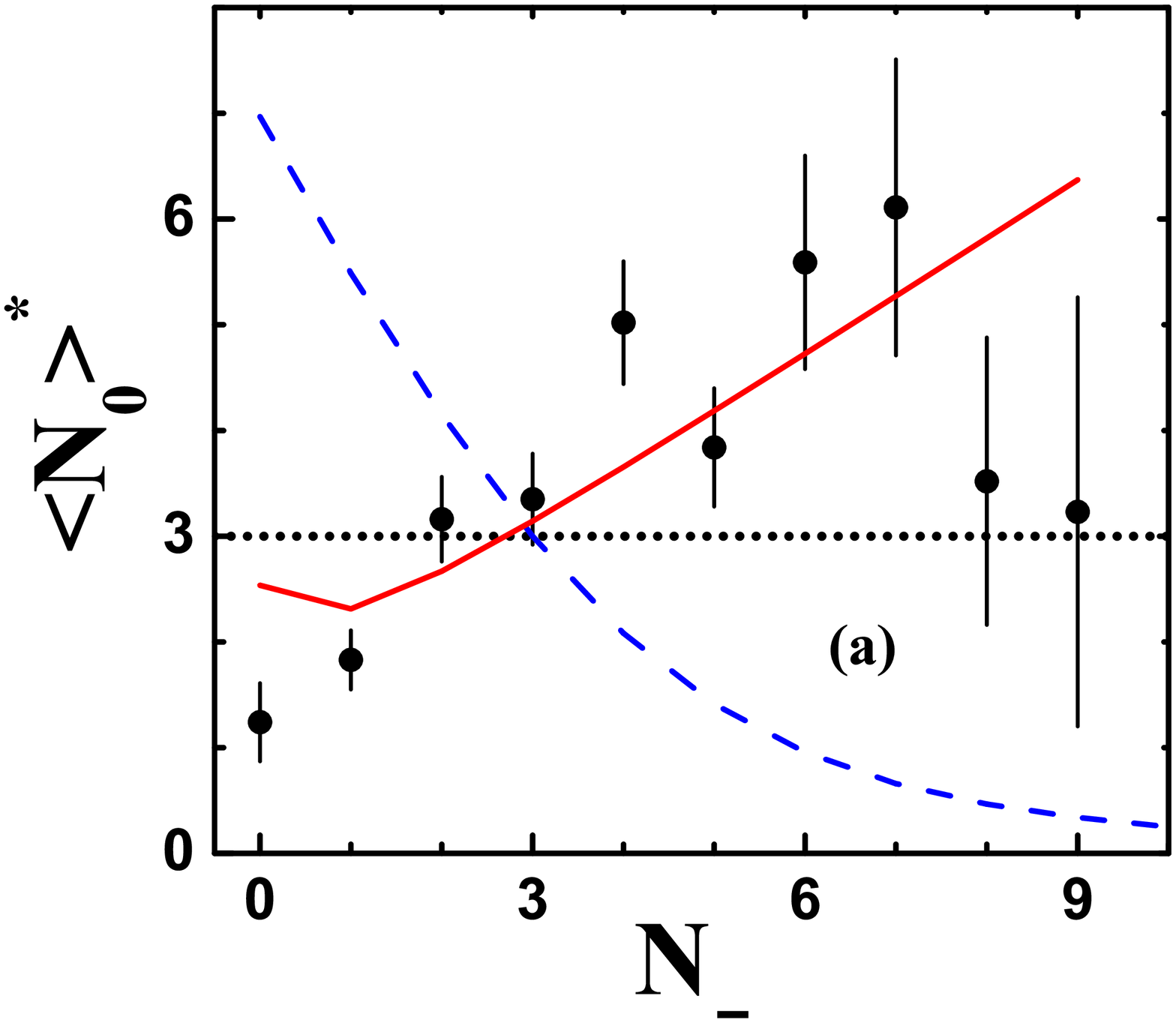,width=0.48\textwidth}\hspace{0.2cm}
 \epsfig{file=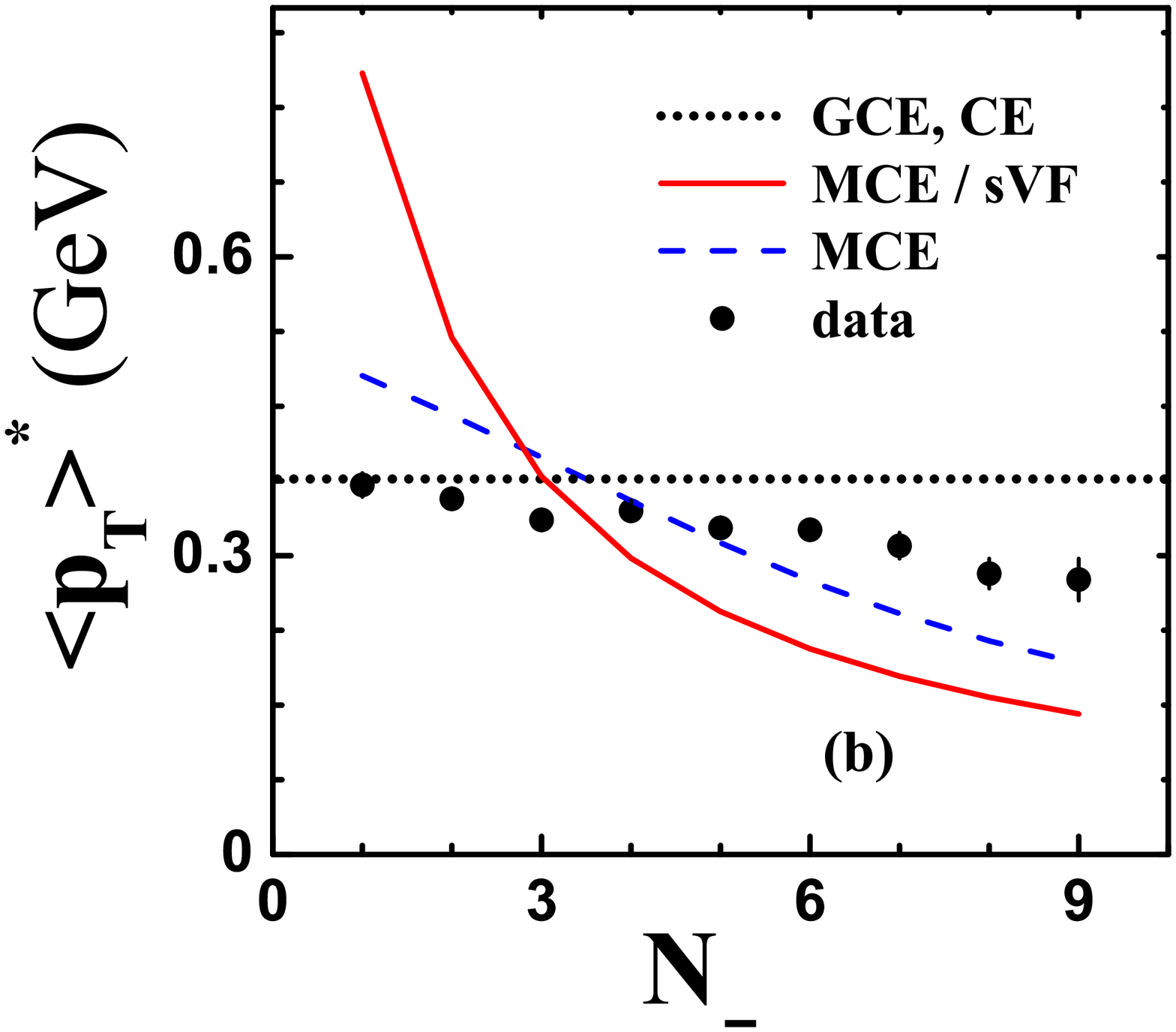,width=0.48\textwidth}
 \end{center}
 \vspace{-0.7cm}
 \caption{(Color online)
The mean multiplicity of neutral particles ({\it left}) and
 the transverse momentum of negatively charged particles ({\it right})
as a function of the multiplicity of negatively charged particles.
The experimental data on $p+p$ interactions  at 205~GeV/c
\cite{Jaeger}, {\it left} and \cite{Kafka}, {\it right} are
indicated by closed circles.  The predictions of the GCE, CE, MCE
and MCE/sVF are shown  by the lines. The calculations are
performed assuming $\overline{N}=3$ and $T=160$~MeV (see text for
details).
 }\label{fig-pT}
\end{figure}

For the purpose of the comparison between considered models and
data the mean transverse momentum was calculated as:
 \eq{
 \langle p_T \rangle^* \;&=\;
        2\int_0^{\infty}dp_T~ p_T^2\int_{-\infty}^{\infty} dy
        ~p_T\;F^*(p)\;\nonumber \\
&\cong ~  \frac{1}{T^{*3}}~\int_0^{\infty}dp_T~ p_T^2\int_{-\infty}^{\infty} dy~
        p_T~\exp\left(-~\frac{p_T\cosh y}{T^*}\right)
  \;=\; \frac{3\pi}{4}\,T^* \;\cong\; 2.36\,T^*\;.
 }
The mean multiplicity of neutral particles calculated within the
models was identified with the mean  $\pi^0$ multiplicity.

The model predictions are summarized in Fig.~\ref{fig-pT}. In the
GCE and CE $\langle N_0\rangle^*$ and $\langle p_T\rangle^*$ are
independent of $N_-$. The MCE reproduces property (B), but leads
to decrease of mean  multiplicity $\langle N_0\rangle^*$ with
increasing $N_-$. Both features (A) and (B) are qualitatively
reproduced in the MCE/sVF.
%The quantitative
%agreement is not very good, especially in Fig.~\ref{fig-pT} right,
%however, one can see that volume fluctuations lead to correlations
%seen also in the data.

%
%%%%%%%%%%%%%%%%%%%%%%%%%%%%%%%%%%%%%%%%%%%%%%%%%%%%%%%%%%%%%%%%%%%%%%%%%%%
%
\section{Summary}\label{S-VI}
Semi-inclusive distributions for the system of neutral and charged
massless particles with net charge equal to zero are considered in
the grand canonical, canonical, micro-canonical ensembles, and in
the micro-canonical ensemble with scaling volume fluctuations
(MCE/sVF). The MCE/sVF has been included in the present study as
it is the only statistical ensemble which reproduces the KNO
scaling of multiplicity distributions and the power law behavior
of the inclusive transverse momentum spectra measured in $p+p$
interactions. The mean multiplicity of neutral particles and
momentum spectra of charged particles are calculated at fixed
charged particle multiplicity $N_-$. Different statistical
ensembles lead to qualitatively different results for these
semi-inclusive quantities even in the large volume limit. In other
words, the semi-inclusive quantities can be different in different
statistical ensembles despite of the ensemble thermodynamical
equivalence.

The obtained model predictions are compared with the experimental
data on $p+p$ inelastic interactions at high energies. The MCE/sVF
follows the trends observed in the data. This demonstrates the
role of volume fluctuations in the system with exact energy and
charge conservation. However, the detailed comparison with the
experimental data is far beyond the scope of this paper.  The
conclusive comparison with the experimental results would require
inclusion in the statistical model calculations several neglected
effects. In particular, the hadron masses and quantum numbers,
isospin symmetry, quantum statistics, and resonance decays should
be taken into account.

%%%%%%%%%%%%%%%%%%%%%%%%%%%%%%%%%%%%%%%%%%%%%%%%%%%%%%%%%%%%%%%%%%%%%%%%%%%
%

\vspace{0.5cm} {\bf Acknowledgments} %---------------------

We would like to thank W. Greiner and M. Hauer for useful
discussions. This work was in part supported by the Program of
Fundamental Researches of the Department of Physics and Astronomy
of National Academy of Sciences, Ukraine. V.V. Begun would like
also to thank for the support of The International Association for
the Promotion of Cooperation with Scientists from the New
Independent states of the Former Soviet Union (INTAS), Ref. Nr.
06-1000014-6454 and the Alexander von Humboldt Foundation for the
support.

%\newpage
\appendix
\section{}\label{app-A}

The bivariate normal approximation
(\ref{PmceG})
of $P_{mce}(N_0, N_-)$
can be derived as follows.
Eq.~(\ref{Omega}) can be rewritten as:
 \eq{
 \Omega_{N_0,N_-}(E,V)
 %\;=\; \frac{1}{E}\;\frac{1}{N!}\; \frac{1}{(N_-)!^2}\;
 %       \frac{A^{N+2N_-}}{(3N+6N_--1)!}
  \;=\; \frac{1}{E}~\exp[f(N_0,N_-)]\;,
 }
where
 \eq{\label{fN}
 f(N_0,N_-)
% &\;=\; \ln[\Omega_{N,2N_-}(E,\,V)]
%         \\
 = (N_0 + 2N_-)\ln[A] \;-\; \ln[N_0!]
  - 2\ln[N_-!] - \ln[(3N_0+6N_--1)!]~,
 }
and $A=VE^3/\pi^2$.
Using the Stirling formula,
$\ln(N!)\cong (N+1/2)\ln(N)-N+\ln(2\pi)/2$
at $N\gg 1$, the r.h.s. of
Eq.~(\ref{fN}) can be expanded with respect to $N_0$ and $N_-$ near the maximum of $ f $.
Then, the mean multiplicities can be calculated from the
condition $\partial f/\partial N_0 =\partial f/\partial N_-=0$.
Second derivatives of $f$ with respect of $N_0$ and $N_-$
at the point of maximum are:
\eq{
 \frac{\partial^2 f}{\partial N_0^2}
 \;\cong\;
% -\frac{1}{N} \;-\; \frac{3}{N + 2N_-}
 % \;\simeq\;
-~\frac{2}{\overline{N}}\;,~~~~
         \frac{\partial^2 f}{\partial N_-^2}
 \;\cong\;
% \;-\; \frac{2}{N_-}
%  \;-\; \frac{12}{N + 2N_-}
 % \;\simeq\;
-~\frac{6}{\overline{N}}\;,~~~~
         \frac{\partial^2 f}{\partial N_0\, \partial N_-}
 \;\cong\; %\;-\;\frac{6}{N + 2N_-}
 % \;\simeq\;
-~\frac{2}{\overline{N}}\;.
 }
and  Eq.~(\ref{PmceG}) follows.

\section{}\label{app-B}
Using the approximation (\ref{PmceG})
the integration over $y$ can be
done analytically. Then the joint $N_0$ and $N_-$
distribution in the MCE/sVF reads:
\eq{\label{PaC}
& P_{\alpha}(N_0,N_-)
 ~\equiv~\int_0^{\infty}dy~P_{mce}(N_0,N_-)~\psi_{\alpha}(y)
 \cong~
%\int_0^{\infty}dy~P^G_{mce}(N,N_-)~\psi_{\alpha}(y)\\
  \frac{k^k}{\Gamma(k)}\,\frac{2\sqrt{2}}{\pi\,\overline{N}}
        \left(\frac{N_0^2+2N_0N_-+3N_-^2}
        {6\overline{N}^2+\overline{N}\,k}\right)^{(k-1)/2}\nonumber \\
&\times~\exp\left(4N_0 + 8N_-\right)~
        K_{1-k}\left[2\sqrt{\left(6+\frac{k}{\overline{N}}\right)
        \left(N_0^2 + 2N_0N_-+3N_-^2\right)}\right]\;,
 }
where $K_{1-k}$ is the Bessel function of the second kind.
Eq.~(\ref{PaC}) can be simplified using the asymptotic
expansion:
 \eq{\label{K-1-k}
 K_{1-k}(x)
 \;=\; \sqrt{\frac{\pi}{2\,x}}\;e^{-x}
       \left(1\;+\;\frac{4k^2-8k+3}{8}\;\frac{1}{x}\;+\;O(x^{-2})\right)\;,
       \quad x\gg 1\;.
 }
Consequently,
 \eq{\label{PaC1}
&P_{\alpha}(N_0,N_-)
 \;\simeq  \; \frac{k^k}{\Gamma(k)}\,\frac{1}{\sqrt{\pi}\,\overline{N}}
            \left(\frac{N_0^2+2N_0N_-+3N_-^2}
            {6\overline{N}^2+\overline{N}\,k}\right)^{(k-1)/2}
        \nonumber    \\
 & \times~  \frac{1}{\sqrt{N_0+2N_-}}~\exp\left[-\,k\,\frac{N_0+2N_-}{3\overline{N}}
            \;-\; \frac{(N_0-N_-)^2}{3N_-} \right].
 }
%
%The MCE/sVF distribution (\ref{Pa-NNpm}) is shown in the
%Fig.~\ref{fig-NNm}, left.
%

\section{}\label{app-C}
In the MCE with a fixed multiplicity $N_-$ the system temperature
$T_{mce}^*$ can be found  as follows. The mean multiplicity of
neutral particles equals to $\langle N^*_0\rangle_{mce}
=VT_{mce}^{*3}/\pi^2$, and their average energy is $\langle
E_0^*\rangle_{mce}=3T_{mce}^*\langle N^*_0\rangle_{mce}$. Thus,
the total energy reads,
 \eq{\label{T*}
 \frac{3VT_{mce}^{*\,4}}{\pi^2} \;+\; 6\,N_-\, T_{mce}^* \;=\; E~.
 }
The first term in the l.h.s. of Eq.~(\ref{T*}) corresponds to the
average energy of neutral particles and the second term to that of
charged particles. For the multiplicity of negatively charged
particles close to the mean multiplicity one can solve
approximately Eq.~(\ref{T*}) with respect to temperature. Denoting
$\delta N_-=N_- -\overline{N}$ and $\delta T= T-T_{mce}^*$ the
solution reads:
 \eq{\label{T1}
 \delta T ~\cong~ -~T~\frac{\delta N_-}{2\overline{N}~+~\overline{N}}~,
 }
or
 \eq{\label{T2}
 T_{mce}^*\;\cong \; T\left(\frac{4}{3}\;-\;\frac{N_-}{3\overline{N}} \right).
 }
Consequently, one gets:
\eq{\label{N0*mce}
\langle
N_0^*\rangle_{mce}~=~\frac{1}{\pi^2}~VT_{mce}^{*3}~\cong~\overline{N}~
\left(\frac{4}{3}\;-\;\frac{N_-}{3\overline{N}} \right)^3~.
}
In the MCE/sVF at fixed $N_-$ one finds,
\eq{\label{T*a}
 3T_{\alpha}^*~\langle N_0^*\rangle_{\alpha}\;+\; 6\,N_-\, T_{mce}^* \;=\; E~.
 }
Using Eq.~(\ref{N0a}) this gives,
\eq{\label{T*a1}
T_{\alpha}^*~\cong~ T~\frac{\overline{N}}{{N_-}}~.
}
%%%%%%%%%%%%%%%%%%%%%%%%%%%%%%%%%%%%%%%%%%%%%%%%%%%%%%%%%%%%%%%%%%%%%%%%%%%%%

\end{document}